%% file: paper.tex
\documentclass[hyperref,onecolumn,apj]{emulateapj}
\usepackage{graphicx}
\usepackage{amsmath}
\usepackage{color}

\newcommand\editremark[1]{{\color{red} #1}}
\newcommand\optional[1]{}

\newcommand\chooseBorC[2]{#2}

\newcommand {\permyr}{${\rm Myr}^{-1}$}

\newcommand{\fb}{f_{\rm b}}
\newcommand{\fbeff}{f_{\rm b,eff}}

\newcommand{\pdf}{{\cal P}}

\newcommand{\PSmoreconstraints}{PSC}
\newcommand{\KimProceedings}{KKL06}
\newcommand{\KimKalogeraLorimer}{KKL}

\newcommand\unit[1]{{\rm #1}}

\shorttitle{BINARY PULSAR BIRTHRATES WITH $\rho(P_{\rm s})$}
\shortauthors{R. O'Shaughnessy and C. Kim}

\begin{document}

\title{PULSAR BINARY BIRTHRATES WITH SPIN-OPENING ANGLE CORRELATIONS}

\author{Richard O'Shaughnessy\altaffilmark{1} AND Chunglee Kim\altaffilmark{2}}
\affil{$^{1}$ 
 Center for Gravitational Wave Physics, Penn State University,
 University Park, PA 16802, USA \\
 $^{2}$ Lund Observatory, Box 43, SE-221 00 Lund, Sweden\\ 
oshaughn@gravity.psu.edu, ckim@astro.lu.se}

\begin{abstract}
One ingredient in an empirical birthrate estimate for pulsar binaries is the fraction of sky subtended by the pulsar
beam: the pulsar beaming fraction.   This fraction depends on both the pulsar's opening angle and the  misalignment
angle between its spin and magnetic axes.  
The current estimates for pulsar binary birthrates are based on an average value of beaming fractions for only two
pulsars, i.e. PSRs B1913+16 and B1534+12. In this paper we revisit the observed pulsar binaries to examine the
sensitivity of birthrate predictions to different assumptions regarding opening angle and alignment.  Based on
  empirical estimates for the relative likelihood of different  beam half-opening angles  and   misalignment angles
  between the pulsar rotation and magnetic axes, we calculate an effective beaming correction factor, $\fbeff$,
  whose reciprocal is equivalent to the average fraction of all randomly-selected pulsars that point toward us.   
 For those pulsars without any direct beam geometry constraints, we find that $\fbeff$ is likely to be smaller
  than 6, a canonically adopted value when calculating birthrates of Galactic pulsar binaries.  We calculate $\fbeff$
  for PSRs J0737-3039A and J1141-6545, applying the currently available constraints for their beam geometry.  As in previous estimates of the
    posterior probability density function P(${\cal R}$) for pulsar binary birthrates ${\cal R}$ ,  PSRs
  J0737-3039A and J1141-6545 still significantly contribute to , if not dominate, the Galactic birthrate of tight
  pulsar-neutron star (NS) and pulsar-white dwarf (WD) binaries, respectively.
 Our median posterior present-day birthrate
  predictions for tight PSR-NS binaries, wide PSR-NS binaries , and tight PSR-WD binaries given a preferred pulsar
  population model and beaming geometry are 89 \permyr,
0.5 \permyr, 
and 34  \permyr, respectively.  For
  long-lived PSR-NS binaries, these estimates include a weak ($\times 1.6$) correction for slowly decaying star formation in the galactic disk. 
For pulsars with spin
  period between 10 ms and 100 ms, where few measurements of misalignment and opening angle provide a sound basis for extrapolation, we
  marginalized our posterior birthrate distribution P(${\cal R}$) over a range of plausible beaming correction factors.
 We explore several alternative beaming geometry distributions, demonstrating our predictions are robust except in (untestable) scenarios with many highly aligned recycled pulsars.   Finally, in addition to exploring alternative beam geometries,  we also briefly summarize how  uncertainties in each pulsar binary's lifetime and in the pulsar luminosity distribution can be propagated into P(${\cal R})$.
\end{abstract}

\maketitle

\section{Introduction}

Using  pulsar survey selection effects to extrapolate outward to the entire Milky Way, the observed  sample of Milky Way
field binary pulsars  constrains the present-day population and birthrate of these binaries, e.g.,
\citet{narayanetal1991}, \citet{phinney1991}, \citet{CurranLorimer1995}, \citet{knst}, \citet{kkl},  henceforth denoted
\KimKalogeraLorimer{}, and references therein.    Along with the properties of the population, this  empirical birthrate
informs models for their formation, e.g., \citet{PSmoreconstraints}  (hereafter PSC), \citet{PSellipticals}; detection
rate estimates for gravitational-wave observatories like LIGO and VIRGO,  \nocite{S3inspiral2008} e.g.\ Abbott et al.\ (2008); and even attempts to unify compact mergers with short $\gamma$-ray bursts \citep{NakarReviewArticle2006}.
Following \KimKalogeraLorimer{}, a posterior prediction for the present-day birthrate (${\cal R}$) of  pulsar binaries on similar
evolutionary tracks to a known pulsar binary can be
expressed in terms of  the pulsar's beaming geometry (through the effective beaming correction factor $f_{\rm b,eff}$), effective lifetime $\tau_{\rm eff}$,  and the population distribution of individual pulsars  (in luminosity and galaxy position, via $N_{psr}$):
\begin{equation}
\label{eq:Posterior:Rate:KnownGeometry}
\pdf({\cal R}) = ({\tau_{\rm eff}}/{N_{\rm psr} f_{\rm b,eff}}){\cal R} e^{-({\tau_{\rm eff}}/{N_{\rm psr}f_{\rm b,eff}}){\cal R}}  \equiv C {\cal R} e^{-C {\cal R}}
\end{equation}
Summing over the individual contributions ${\cal R}_i$ from each specific pulsar binary $i$, a posterior prediction for
the  overall  Galactic birthrate is
\begin{equation}
{\cal P}({\cal R}_{tot}) = \int \Pi~ d{\cal R}_i {\cal P}_i({\cal R}_i)  ~\delta ({\cal R}_{tot} - \sum {\cal R}_i) ~.
\end{equation}
%
As of 2009, the best constrained $f_{\rm b}$'s for binary pulsars are available for PSRs B1913+16 and B1534+12 \citep{knst}. Previous works taking an empirical approach relied on these two pulsars for the beaming correction to the rate estimates, e.g. \KimKalogeraLorimer{}, \citet{kalogera04}, \citet{kkl06} (hereafter \KimProceedings{}). The average value of $f_{\rm b}\sim6$  based on PSRs B1913+16 and B1534+12 was used as `canonical' value in order to calculate the birthrate (or merger rate) of pulsar binaries and the inferred detection rates for the gravitational-wave detectors. 

The motivation for this paper is to provide not only updated  Galactic birthrates of pulsar binaries, but also to
  provide and explain  more generic
  beaming correction factors for use in the birthrate estimates.
In this work, we introduce an empirically-motivated beaming model, derive a probability distribution function for $f_{\rm b}$,
and calculate the effective beaming correction factor $f_{\rm b,eff}$ for two types of pulsar binaries, a pulsar with a 
  neutron star (PSR-NS) or a white dwarf (PSR-WD) companion. Specifically, we adopt currently available
constraints on a misalignment angle $\alpha$ between pulsar spin and magnetic axes, e.g., \citet{GilHan96},
\citet{2003PASJ...55..461Z}, \citet{kol04}, as well as the empirical relationship betwen the half-opening angle $\rho$
and pulsar spin period $P_s$ as priors (e.g. Kramer et al.\ 1998).  For mildly recycled pulsars ($10~{\rm ms} < P_s <
100~{\rm ms}$), we  anchor our theoretical expectations with observational bias; see \S{\ref{sec:rateconst}} for details.

\section{Rate constraints for pulsar binaries}
\label{sec:rateconst}

Following  \KimKalogeraLorimer{},  we assume each  pulsar $i$ in a pulsar binary is equally visible throughout its effective lifetime $\tau_i$ everywhere  along a fraction   $f_{b,i}$ of all lines of sight from the pulsar, ignoring any 
 bias introduced through time evolution of  luminosity and opening angle.\footnote{%
 Pulsar spin precession occurs on a much shorter timescale and does not violate our assumption, if $f_{b,i}$
correctly accounts for the precession-enhanced extent of the emission cone.  }
Consistent with observations of the Milky Way's star formation history \citep{2001ASPC..230....3G},
we also assume the star formation rate and pulsar birthrate is constant over the past $\tau_{mw}=10 ~\unit{Gyr}$. 
Based on these two approximations, the mean number $\bar{N}_i$ of pulsars on the same  evolutionary track as $i$ that are visible at
present can be related to its birthrate ${\cal R}_i$ by 
\begin{eqnarray}
\label{eq:MeanNumberViaRate:KnownGeometry}
\bar{N}_i = f_{\rm b}^{-1} {\cal R}_i \tau_i
\end{eqnarray}
Conversely, Eq.\ (\ref{eq:MeanNumberViaRate:KnownGeometry}) and Bayesian Poisson statistics  
uniquely determine our posterior prediction for each pulsar's birthrate shown in Eq.\  (\ref{eq:Posterior:Rate:KnownGeometry}).

\subsection{Number of pulsars}
When inverting relation \ref{eq:MeanNumberViaRate:KnownGeometry} to calculate a birthrate ${\cal R}_i$, we determine
$N_{\rm psr}$ following the procedure originally described in KKL.  In order to examine the effects of newly
discovered pulsars and estimated pulsar beaming fractions, we considered the same surveys we used in
\PSmoreconstraints.  This includes all surveys listed in KKL and three more surveys, i.e. the Swinburne
intermediate-latitude survey using the Parkes multibeam system \citep{swinburne}, the Parkes high-latitude survey
\citep{J0737Adiscovery}, and the mid-latitude drift-scan survey with the Arecibo telescope \cite{J1824discovery}.
Explicitly,  $N_{\rm psr}$ is the most likely number of a given pulsar binary population in our Galaxy.  We estimate
$N_{psr}$ via a synthetic survey of  $10^6$ similar pulsars pointing towards us, as described in \KimKalogeraLorimer.  If $N_{det}$ synthetic pulsars are
found in our virtual survey,  we estimate  $N_{psr}  = 10^6/N_{det}$; this ratio agrees with the slope labelled $\alpha$ in
\KimKalogeraLorimer{} and defined in their Eq. (8).   Within  Poisson error of
a few \% (=$1/\sqrt{N_{det}}$), our results are consistent with a reanalysis of previous simulations (cf. Table 1 in
\PSmoreconstraints, from \KimProceedings,\KimKalogeraLorimer).

\subsection{\label{sec:sub:lifetime}Effective lifetime}

In the relation \ref{eq:MeanNumberViaRate:KnownGeometry}, the \emph{effective} lifetime $\tau_i$ encapulsates any and all factors needed to convert between  the present-day 
number $\bar{N}_i$ and a birthrate $f_b^{-1}{\cal R}$ of pulsars that emit along our line of sight.
For example, pulsars with a long visible lifetime $\tau$ do not reach their equilibrium number (${\cal R} \tau$), instead
accumulating steadily with time.  
Assuming a steady birthrate, the {\it effective} lifetime for a binary contaning a pulsar $i$ is the smaller of the
Milky Way's age and the pulsars' lifetime:
\begin{equation}
\label{eq:def:taui}
\tau_i        = \text{min}(\tau_{mw},\tau_{age}+\text{min}(\tau_{mrg},\tau_{d}))
\end{equation}
The lifetime of a pulsar binary is defined as the sum of the current age of the system ($\tau_{age}$) and an estimate of
the remaining \emph{detectable} lifetime.  For the present age of PSRs B1913+16 and B1534+12, we use the spin-down age  of a pulsar, an estimate based on extrapolating pure magnetic-dipole spindown backwards from present to some  high initial frequency  \citep{Ar}. 
 For all other pulsars, we adopt the proper-motion corrected ages presented in \citet{KT09agecorrection}.
  Because the pulsar spends most of its time at or near
its current age, the error in the current age should be relatively small for any recycled pulsar,  unless its
  current state is very close to the endpoint of recycling.   
For pulsar binaries considered in this work, even  the most extreme possibility  -- a true age $\tau_{age}=0$ -- changes lifetimes by less than 30\%.
 The only exception is PSR J0737-3039A (see \S \ref{sec:Rates:PSRNS:Merging}).
 For sufficiently tight binaries (orbital periods less than $\sim$ 10 hours),
the remaining lifetime is usually limited by inspiral through gravitational radiation \citep{peters64}. 
When the second-born pulsars are observed, however, the binary's  detectable lifetime is instead  limited  by how long it radiates, as measured by the time until it reaches the pulsar ``death-line'' ($\tau_{d}$) as shown in Fig.\ \ref{fig:ppdotDeathlines} 
\citep{ChenRuderman:1993, 2000ApJ...531L.135Z, 2002ApJ...576..366H, psr-theory-spindown-Spitkovsky2005}.   
We note that our results are fairly robust to the uncertainties in the
death-line: (a) merging timescales are more important in birthrate estimates for recycled pulsars ($\tau_{\rm mrg} >
\tau_{\rm d}$), and (b) the death-line is relatively well-determined for the non-recycled pulsars in our sample,
see, e.g., the discussion in \S \ref{sec:sub:Rates:PSRWD} of PSRs J1141-6545 
 and in \S \ref{sec:Rates:PSRNS:Merging} of  J1906+0746 as well as Figure \ref{fig:ppdotDeathlines}.\footnote{We estimate  the death timescale $\tau_{\rm d}$ using \citet{ChenRuderman:1993} (see Table \ref{tab:psrns}, \ref{tab:psrwd} and Fig.\ \ref{fig:ppdotDeathlines}), which is typically the shortest
  among different models presented by  
\citet{2000ApJ...531L.135Z},
  \citet{2002ApJ...576..366H},
\citet{psr-theory-spindown-Spitkovsky2005}. We note that the uncertainty in $\tau_{d}$ for PSR J1141-6545 is roughly 60\%, the
  death timescale for PSR J1141-6545 estimated by a \citet{ChenRuderman:1993} death-line is $\sim$0.10 Gyr, while
  \citet{psr-theory-spindown-Spitkovsky2005} curve predicts $\sim$0.17 Gyr. 
}
%

The  reconstructed birthrate for wide binaries ($\tau_i\simeq O(\tau_{mw})$) is sensitive to variations in the star
formation history of the Milky Way.  Small volumes of the Milky Way, such as the Hipparchos-scale volume of stars,
can have $O(1)$ relative changes in the star-formation history fluctuation on $O(0.3-3~ \unit{Gyr})$ timescales (see for
example Fig.\ 4 in \cite{2001ASPC..230....3G}).  On the larger scales over which these radio pulsar surveys are
sensitive, however, these fluctuations average out; see for example  observations of open clusters and well-mixed dwarf
stars in \cite{2004NewA....9..475D}, \cite{2001MNRAS.327..329H}, and references therein.  In particular, for lifetimes
$\tau_i<0.2~\unit{Gyr}$
relevant to the most significant tight PSR-NS binaries, the star formation rate is constant to within tens of percent,
from Fig.\ $15$ in \cite{2004NewA....9..475D}.
On longer timescales,  observations of other disk galaxies and phenomenological models for galaxy assembly  also support a nearly-constant star formation rate
(see \citet{2006MNRAS.366..899N}, \citet{mw-sfr-SchonBinney2009},  \citet{2009AJ....137..266F} and references therein).
For pulsar binaries with $\tau_{i} > 3 ~\unit{Gyr}$, observations and models suggest the  star
formation rate trends weakly upward with time; see, e.g., 
\cite{mw-LocalStructure-AumerBinney2009} and \cite{2009AJ....137..266F}.

For simplicity and to facilitate comparison with previous results, we  adopt a constant star formation rate in  most of
this paper and figures; 
our final best estimates (Figure \ref{fig:GlobalAmbiguity}) include a small correction for exponentially-decaying disk
star formation, based on 
 the candidate star formation history 
\begin{eqnarray}
\label{eq:MeanNumberViaRate:CandidateSFR}
{\cal G} = \frac{\dot{\Sigma}_{*}(t)}{\dot{\Sigma}_*(0)}&\propto&  e^{- 0.09t/\unit{Gyr}}
\end{eqnarray}
where $t=0$ is the present, drawn from \citet{mw-LocalStructure-AumerBinney2009}; other proposals, such as profiles expected from a
Kennicutt-Schmidt relation \citep{2009AJ....137..266F}, are easily substituted.  Assuming negligible delay between star formation and compact binary formation (i.e., ${\cal R}\propto {\cal G}$), Eq.\ (\ref{eq:MeanNumberViaRate:KnownGeometry}) for the average number of pulsars seen $\bar{N}$ at present in terms of the present-day birthrate ${\cal R}$
generalizes to 
\begin{eqnarray}
\label{eq:MeanNumberViaRate:KnownGeometry:TimeDependent}
\bar{N}_{X, \text{ with time}} = f_{\rm b}^{-1} R_X(0) \int_{-\min(\tau_X,\tau_{mw})}^o  {\cal G} \; dt = N_{X} \tau_{i,X}
\left<{\cal G}\right>
\end{eqnarray}
  Based on the candidate star formation history of Eq. \ref{eq:MeanNumberViaRate:CandidateSFR}, $\left<\cal
    G\right>\approx 1.6$ for the longest averaging time.  Thus, because there was more star formation available to form the
widest, long-lived PSR-NS binaries than if stars formed at a steady rate, the present-day formation rate for these wide PSR-NS binaries is
roughly $1/1.6\approx 0.65$ times smaller than that shown in Fig.\ \ref{fig:Rates:PSRNS:Wide}.


\begin{figure}
\chooseBorC{
\includegraphics[width=\columnwidth]{fig-mma-DeathLines-BW}
\caption{\label{fig:ppdotDeathlines} $P_{\rm s}-\dot{P_{\rm s}}$ diagram superimposed with various death-lines.  The
  stars indicate binaries whose visible lifetime is dominated by the death timescale $\tau_{\rm d}$: all wide PSR-NS
  binaries (recycled) and PSRs J1141-6545 and J1906+0746 (nonrecycled). The thick solid line is the Chen and Ruderman
  death-line \citep{ChenRuderman:1993}; the dotted lines are death lines from
  \citet{psr-theory-spindown-Spitkovsky2005} and \citet{2000ApJ...531L.135Z}. The thin solid line is the   curve ${P_{\rm s}}/{2{\dot P}_{\rm s}}=10 \rm {Gyr}$; the   dashed line is the curve where ${\dot E} =4\pi^{2} I {\dot P}{_s} P{_s}^{-3}=10^{30}$ erg/s, assuming a moment of inertia I=10$^{45}$ g cm$^{2}$
  \citep{LorimerKramerPulsarBook}.  
 }
}{
\includegraphics[width=\columnwidth]{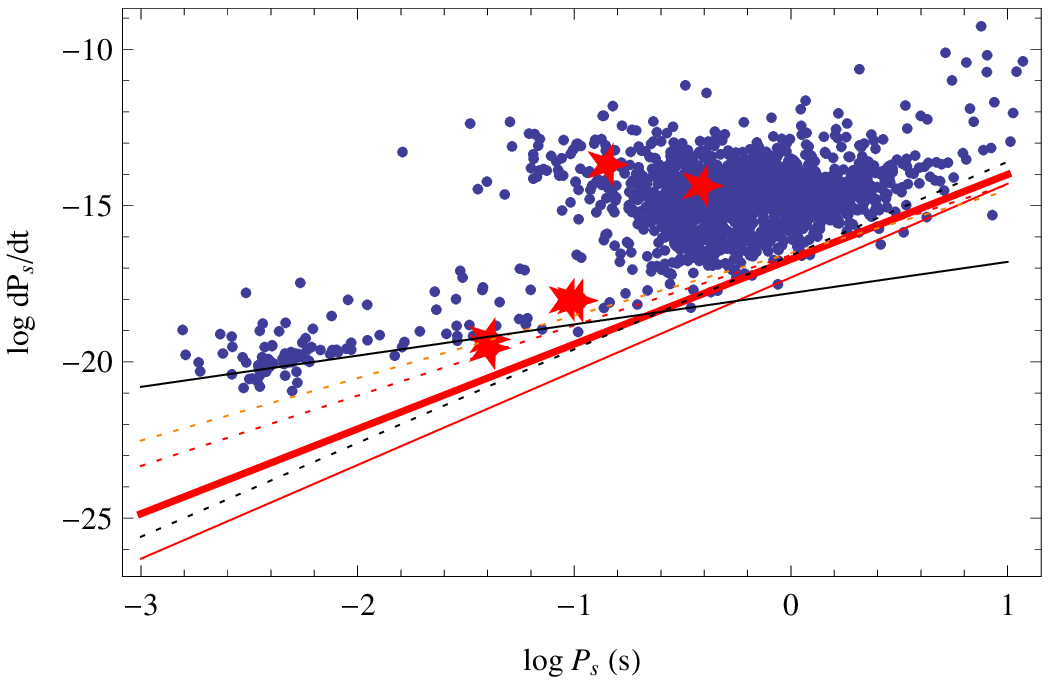}
\caption{\label{fig:ppdotDeathlines} 
$P_{\rm s}-\dot{P_{\rm s}}$ diagram superimposed with various death-lines.  The stars indicate binaries whose visible lifetime is dominated by the death timescale: all wide PSR-NS binaries (recycled) and PSRs J1141-6545 and J1906+0746 (nonrecycled). The thick solid line is the Chen and Ruderman death-line \citep{ChenRuderman:1993}; the thin solid line is from \citet{psr-theory-spindown-Spitkovsky2005}; the dotted red lines are  from \citet{2000ApJ...531L.135Z}. The black solid line is the   curve ${P_{\rm s}}/{2{\dot P}_{\rm s}}=10 \rm {Gyr}$; the  black dotted line is the curve where ${\dot E} =4\pi^{2} I {\dot P}{_s} P{_s}^{-3}=10^{30}$ erg/s, assuming a moment of inertia I=10$^{45}$ g cm$^{2}$ \citep{LorimerKramerPulsarBook}. 
 }
}  
\end{figure}


\subsection{Beaming Distribution}
\label{sec:sub:beaming}


Except for two pulsars,  the  empirical beaming geometry of any pulsar in a binary is not
tightly constrained.  Further, as discussed below, observations of single pulsars suggest that, even
restricting to  pulsars with similar evolutionary state (e.g., spin), the pulsar beam is randomly aligned relative to
its spin axis.
Nonetheless, because of the poisson statistics of pulsar detection (\KimKalogeraLorimer{}), only one property of the intrinsic pulsar geometry
distribution matters: the  fraction $F(P_s) [\equiv f_{\rm b,eff}^{-1}]$  of all randomly selected pulsars whose beam crosses our line of sight.  
Assuming the pulsar beam
  drops off rapidly -- typical models involve gaussian cones --   the fraction of all pulsars $F(P_s)$ of a given spin
  period and luminosity that emit towards us is well-defined and essentially independent of distance.   The main
  ingredients necessary to calculate $F(P_s)$ are the half-opening angle $\rho$ of the radio beam and the 
misalignment
  angle between pulsar's spin and magnetic axes $\alpha$.  
%
 All of our analysis leading to the posterior rate prediction, Eqs.\ (\ref{eq:Posterior:Rate:KnownGeometry}) $-$ (\ref{eq:MeanNumberViaRate:KnownGeometry}) generalize to an arbitrary beam geometry  distribution upon substituting $1/f_b\rightarrow F$. 


To calculate $F$, for simplicity and following historical convention we assume the bipolar pulsar beam subtends a hard-edged cone with opening
angle $\rho$, misaligned by $\alpha<\pi/2$ from the rotation axis.  As the pulsar rotates, this beam subtends a solid angle
\begin{eqnarray}
\label{eq:def:fb}
\frac{4\pi}{f_{\rm b}}& =& {2\pi}\times 2 \int_{\rm {min}(0,  \alpha-\rho)}^{\rm {max}(\alpha+\rho,\pi/2)} d\cos\theta~,
\end{eqnarray}
where the beaming correction factor $f_{\rm b}$ is the inverse of the fraction of solid angle the beam subtends{, given
  opening and inclination angles ($\rho,\alpha$)}. This simple model for beam geometry and its correlation with pulsar
spin have been explored ever since pulsar polarization data and the rotating vector model made alignment constraints
possible, see, e.g., \citet{1993ApJS...85..145R}, \citet{GilHan96}, \citet{krameretal98},
\citet{WJ08interpulses}.
In one form or another, this theoretically-motivated semi-empirical correlation has been adopted as an ingredient in most models for
synthetic pulsar populations: see, e.g., \citet{Ar}, \citet{2004ApJ...604..775G}, \citet{2006ChJAS...6b..97G}, \citet{2006ApJ...643..332F}, \citet{2007ApJ...671..713S}, and references therein.\footnote{Though some authors also adopt an
   orientation-dependent flux based on classical core/cone models for pulsar emission, see \citet{Ar}, \cite{2007ApJ...671..713S} and references therein,  in order to illustrate the influence of spin-dependent beaming on rate predictions, we assume uniform emission over a cone.}   
Conversely, fully self-consistent population models which account for space distribution and kicks, luminosity evolution,  beam structure and shape evolution, accretion during binary evolution, and spindown are required, in order to extract all the degenerate parameters that enter into a model by comparing its predictions with observations. 
%
%

In this paper, we focus on quantifying how much the addition of a spin-dependent opening angle $\rho(P_{\rm s})$  
  combined with a misalignment angle $\alpha$ distribution influences birthrate estimates. As a simple approximation, 
%
we adopt $\alpha$ and $\rho(P_{\rm s})$ distributions that  are consistent with the observed pulse widths at 10\%
intensity level; see \citet{2003PASJ...55..461Z} and \citet{kol04}.  
Specifically, we assume the misalignment angle is uniformly distributed  between [0,$2/\pi$]:
\begin{eqnarray}\label{eq:alphaflat} 
\pdf(\alpha)d \alpha = 2/\pi d\alpha~.
\end{eqnarray}
Other distributions for $\alpha$ have been proposed, from a random vector $\pdf(\alpha)=\sin \alpha$ (strongly disfavored by the high
  frequency with which low-$\alpha$ pulsars have been detected; most detected pulsars would be orthogonal) to more
  tightly aligned distributions; see  for example Fig.\ 3 in \citet{kol04}.  While limited current observations cannot tightly
  constrain the intrinsic misalignment angle distribution, they strongly suggest tight alignments (low $\alpha$) should
  be attained at least as often as a flat distribution implies.  
   In Table \ref{tab:Conclusions} we compare our reference model
  with several alternative misalignment distributions; only for exceptional misalignment assumptions (e.g.,
  $p(\alpha)\propto 1/\alpha$) will our final results change significantly.

For pulsars with spin period larger than $10 \unit{ms}$, we calculate $\rho(P_{\rm s})$ applying a model 
consistent with classical observations of isolated pulsars's beam geometry \citep[see,e.g.,][and references therein]{WJ08interpulses}, as shown in  Fig.\ \ref{fig:Ingredients:SinglePulsarOpeningAngles}
\begin{subequations}
\label{eq:RhoOfP}
\begin{eqnarray}
\rho(P_{\rm s})&=&
5.4^o P_s^{-0.5} \qquad P_s>10~\unit{ms} 
\end{eqnarray}
The available single-pulsar  data does not support a compelling model for rapidly spinning young and recycled pulsars.
We adopt an ad-hoc power-law form as our fiducial choice:\footnote{The short-period extrapolation used
  here has no practical impact on our results; see Footnote 6.  For example, we have also considered the much tighter beams
  implied by $\rho\propto P^{2}$; except for a handful of still-irrelevant PSR-WD binaries, our birthrate estimates are unchanged. }
\begin{eqnarray}
&=& 54^o(P_s/10\unit{ms})  \qquad P_s<10~\unit{ms}
\end{eqnarray}
\end{subequations}
In both regions, we allow for a  small gaussian error in $\ln \rho$ to allow for model uncertainty; 
we conservatively adopt $\sigma_{\ln\rho}=\ln 1.3$ (i.e., ``30\%'' error; Figure
\ref{fig:Ingredients:SinglePulsarOpeningAngles} shows a $2\sigma$ interval about our fiducial choice).\footnote{Though the apparent beam size depends on the observing frequency slightly, because our standard error is typically much larger than the change due to observing frequency, if
  any, we also adopt a  frequency-independent emission cone;
   cf. the discussion in \citet{2002ApJ...577..322M} and \cite{2008MNRAS.388..261J}.
}
Based on  these two independent distributions, we estimate the  fraction of pulsars with spin period $P$ that point towards us as
\begin{eqnarray}
\label{eq:def:fbeff}
F(P_{\rm s})&\equiv& \left< f_{\rm b}^{-1} \right>\equiv 1/f_{\rm b,eff} 
\end{eqnarray}
The trend of $f_{\rm b,eff}(P_{\rm s})$ shown in Fig.\ \ref{fig:fig-mma-Models-fbEffVersusP} largely agrees with previous
estimates of the beaming fraction for nonrecycled pulsars; see, e.g.,  \cite{1998MNRAS.298..625T}.  

In addition to our fiducial choice, we have explored several other short-period beaming models.  As a benchmark for comparison, two extreme cases are
provided in Figure \ref{fig:Ingredients:SinglePulsarOpeningAngles}: a ``very narrow''
(\chooseBorC{solid}{blue})
 and ``very wide''  (\chooseBorC{dashed}{green})
opening angle model, where the $P^{-1/2}$
truncates at $14^\circ$ ($149 \unit{ms}$) and $54^\circ$ ($10 \unit{ms}$), respectively.   For relevant spin periods, these extreme alternatives imply noticeably different amounts
of beaming correction from each other and our fiducial model, up to $O(\times 2)$; see Figure
\ref{fig:fig-mma-Models-fbEffVersusP}.    Nonetheless, our best semi-empirical estimates for  binary pulsar birthrates are fairly or highly insensitive to the precise opening
angle model
 $\rho(P)$ adopted for rapidly spinning pulsars.\footnote{For example, while these alternatives can lead to slightly
  different values for $f_{\rm b,eff}$ for  tight PSR-WD binaries, the high birthrate
  and long period of PSR J1141-6545 makes the details of a short-period extrapolation astrophysically
  irrelevant; see \S \ref{sec:sub:Rates:PSRWD}.
  The merger rate of tight PSR-NS binaries is dominated by PSRs J1906+0746 ($P>100\unit{ms}$) and J0737-3039A (limited both
  through constrained
  beam geometry and through anchored expectations from PSRs B1913+16 and B1534+12).  The merger rate of wide PSR-NS binaries is dominated by two pulsars with $P\simeq 100 \unit{ms}$,
  where single-pulsar observations strongly constrain reasonable $\rho$ choices.
}
As described below, for pulsars with $10 \unit{ms} < P_s < 100 \unit{ms}$, we anchor our theoretical bias with observed
geometries of comparable binary pulsars.   In this period interval,  the choice for $\rho(P)$ effectively serves as an
\emph{upper bound} on plausible $\rho$ (and sets a lower bound on $\fb$).


\begin{figure}
\chooseBorC{
 \includegraphics[width=\columnwidth]{fig-mma-paper-OpeningAngle-BW}
\caption{\label{fig:Ingredients:SinglePulsarOpeningAngles} Opening angles $\rho(P_{\rm s})$ versus spin period for a
  sample of pulsars drawn from \citet{krameretal98} (solid circles) and PSRs B1913+16 \citep{WeisbergTaylor02} and
  B1534+12 \citep{Arzoumanian1996B1534}. 
For PSRs B1913+16 and B1534+12, where multiple conflicting measurements of $\rho$ exist, all values are shown (triangles).   
For pulsars with 
P$_{\rm s}>$ 100 ms,  we adopt an empirically-motivated relation for $\rho(P_s)$, 
$\rho \propto$ P$_{\rm s}^{-0.5}. $
  This distribution,  allowing for scatter (the  shaded region) is
  used to calculate the effective beaming correction factor $f_{\rm b,eff}$ over many pulsars with a similar spin; see
  Eq. \ref{eq:def:fbeff}.   For pulsars with spin period 10 ms $<$ P$_{\rm s}$ $<$ 100 ms, we define an alternative
  `conservative' beaming model  such that $f_{b}=6$ (not shown).  In this interval, to reflect our uncertainty in pulsar opening
  angle models, we assume $\ln f_{\rm b}$ is uniformly distributed between the `standard' beaming model implied by this
  figure's $\rho(P)$ and the `conservative' choice $f_b=6$.
 Finally, the thin and dashed lines indicate two extreme alternative opening angle models discussed in \S
 \ref{sec:sub:beaming}.
 }
}{
 \includegraphics[width=\columnwidth]{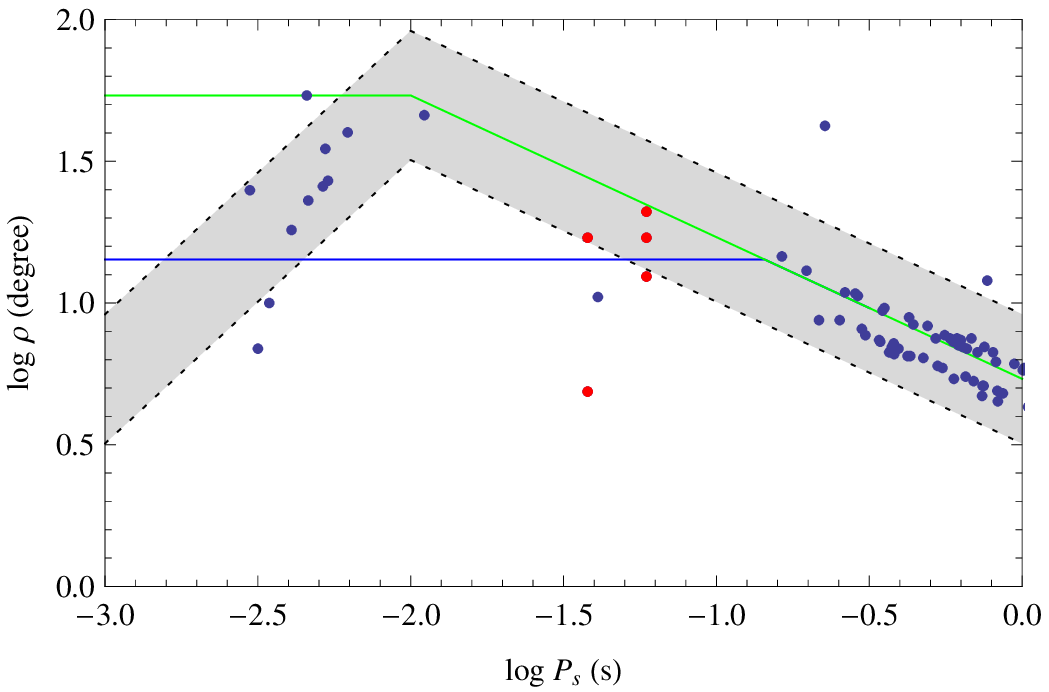}
\caption{\label{fig:Ingredients:SinglePulsarOpeningAngles} Opening angles $\rho(P_{\rm s})$ versus spin period for a sample of pulsars drawn from \citet{krameretal98}  and PSRs B1913+16 \citep{WeisbergTaylor02} and B1534+12 \citep{Arzoumanian1996B1534}. 
For pulsar binaries where multiple
  conflicting measurements of $\rho$  exist, all values are shown (red points).  For pulsars with P$_{\rm s}<$ 10 ms or P$_{\rm s}>$ 100 ms,  we
  adopt an empirically-motivated relation for $\rho(P_s)$, including    $\rho \propto$ P$_{\rm s}^{-0.5}$ at
  large $P$; this distribution,  allowing for scatter (the  shaded region) is
  used to calculate the effective beaming correction factor $f_{\rm b,eff}$ over many pulsars with a similar spin; see
  Eq. \ref{eq:def:fbeff}.   For pulsars with spin period 10 ms $<$ P$_{\rm s}$ $<$ 100 ms, we define an alternative
  `conservative' beaming model  such that $f_{b}=6$ (not shown).  In this interval, to reflect our uncertainty in pulsar opening
  angle models, we assume $\ln f_{\rm b}$ is uniformly distributed between the `standard' beaming model implied by this
  figure's $\rho(P)$ and the `conservative' choice $f_b=6$. 
 Finally, the thin and dashed lines indicate two extreme alternative opening angle models discussed in \S
 \ref{sec:sub:beaming}.
 }
}
\end{figure}

\begin{figure}
\chooseBorC{
 \includegraphics[width=\columnwidth]{fig-mma-paper-fbEffVersusP-BW}
}{
 \includegraphics[width=\columnwidth]{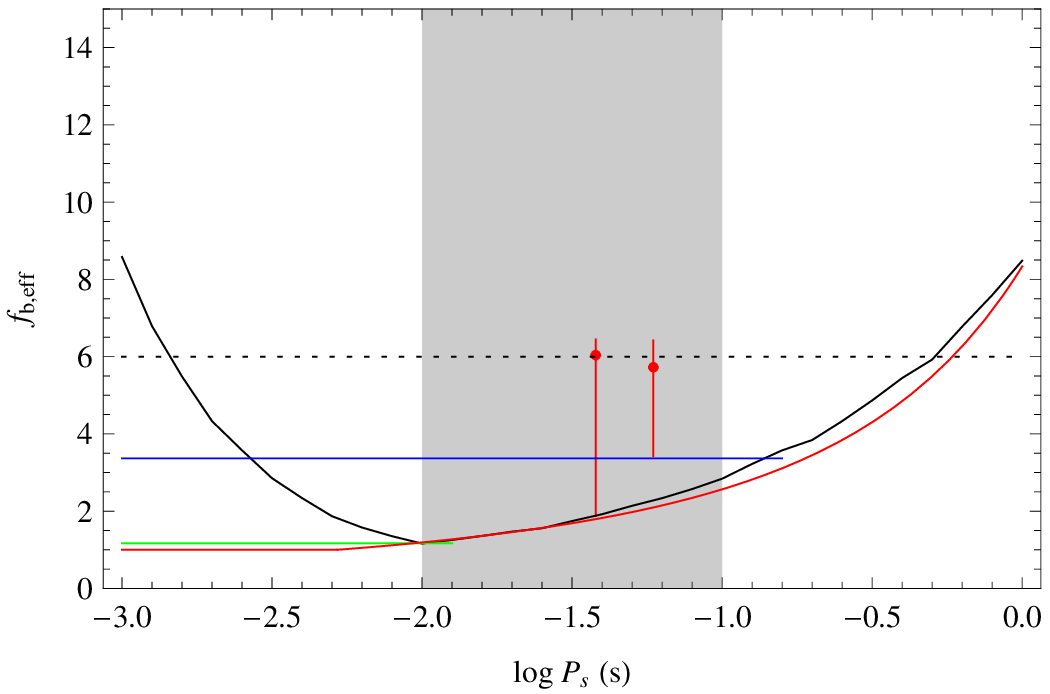}
}
\caption{\label{fig:fig-mma-Models-fbEffVersusP} The  effective beaming correction factor $\fbeff$ versus pulsar spin
  period $P_s$ (\chooseBorC{thick solid}{black}), assuming the distribution of beaming geometries described in the text.  For comparison, we show comparable estimates for the beaming  correction factor from   \citet{1998MNRAS.298..625T}  as a \chooseBorC{dotted}{red}{} curve (using their Eq.\ 15, but extended beyond the $P=0.1-5~\unit{s}$ region they fit)
 as well as the observed $P_s-f_{\rm b}$ combinations for PSRs B1913+16 (log P$_{\rm s}=-1.23$) and B1534+12 (log P$_{\rm s}=-1.42$), based on the preferred values adopted in
 \citet{knst} (\chooseBorC{triangles}{red points}) and all possible combinations of values for $\alpha$ and $\rho$ provided  in \cite{Arzoumanian1996B1534}, \cite{krameretal98} (\chooseBorC{}{ red }vertical bar).
Finally, the gray shaded region indicates the spin period interval  (10 ms $< P_s<$ 100 ms), where few pulsar opening angles are available to
constrain our model; see \citet{krameretal98}.
Also shown  \chooseBorC{as thin dashed and solid lines}{in green and blue}{} are the effective beaming factors implied for the extreme ``narrow'' and
  ``wide'' models shown in Fig. \ref{fig:Ingredients:SinglePulsarOpeningAngles}.
}
\end{figure}


Any specific choice of opening angle and misalignment distribution $\pdf(\alpha,\rho|P_{\rm s})$ determines that model's  probability $\pdf(f_b|P_{\rm s})$ that a
randomly selected pulsar with spin period $P_{\rm s}$ has beaming correction factor $f_b$
\begin{eqnarray}
\label{eq:pdf:Intrinsic}
\pdf(f_b|P_{\rm s})d f_b &=&\int  \delta(f_b-f_b(\alpha,\rho)) \pdf(\rho|P_{\rm s})d\rho \pdf(\alpha)d \alpha 
\end{eqnarray}
Further, because $1/f_b$ is the probability that a given randomly selected pulsar points towards us, the distribution
$p_e$ of
$f_b$ among the fraction  $F$ of all pulsars aligned with our line of sight is
\begin{eqnarray} 
\label{eq:pdf:Observed}
\pdf_{e}(f_b|P_{\rm s})df_b &=&  \frac{\pdf(f_b|P_{\rm s})/f_b}{F(P_{\rm s})}
\end{eqnarray}
For our fiducial beaming model,  Fig.\ \ref{fig:Known:FbVersusRho} compares the median value of $f_b$ for the two distributions as a function of
  $\rho(P_{\rm s})$.
It shows the  beaming correction factors $f_b$ for the detected population of pulsars can differ substantially from the underlying population, if only because those pulsars with narrow
beam coverage (i.e.,
$f_b\gg 1$) are less likely to be detected.
Only extremely narrow pulsar beams $\rho<10^o$ will lead to a typical \emph{detected} pulsar
with $f_b\simeq 6$, comparable to the measured value for known PSR-NS binaries that was adopted in previous analyses as
the canonical value; see Fig.\ \ref{fig:fig-mma-Models-fbEffVersusP}.

\begin{figure}
\chooseBorC{
\includegraphics[width=\columnwidth]{fig-mma-KnownPulsars-VersusOpeningAngle-BW}
\caption{\label{fig:Known:FbVersusRho} The median value of $f_{\rm b}$ for detectable pulsars (triangles;  obtained from $\pdf_e$  in Eq.\ (\ref{eq:pdf:Observed})) and all pulsars
  (points; from $\pdf$  in Eq.\ (\ref{eq:pdf:Intrinsic})) versus $\rho$, assuming  a synthetic pulsar population with misalignment angle $\alpha$  uniformly
  distributed ($p(\alpha)d\alpha = 2d\alpha/\pi$) and beaming described by Eq. (\ref{eq:def:fb}).  
Also shown (solid) is $1/f_{\rm b} = (1-\cos\rho) + (\pi/2 -\rho)\sin\rho$, an estimate of $f_{\rm b}(\rho)$ assuming
misalignment angle $\alpha$
is uniform in \emph{area} ($p(\alpha)d\alpha=\sin \alpha d\alpha$), rather than angle as adopted here; see  Emmering and Chevalier (1989). 
The \emph{detectable} pulsar population will preferentially have wide beams.   Only a population of pulsars with
  very tight opening angles  are consistent with the canonical value of $f_{\rm b}=6$.
 In our ``standard'' model (Eqs.\ \ref{eq:RhoOfP} and \ref{eq:def:fbeff}), such narrow beams are improbable for pulsars 
  spinning with   $10 \unit{ms}<P_s<100\unit{ms}$ that is relevant to most of the known PSR-NS binaries (see Table \ref{tab:psrns}).
 }
}{ 
\includegraphics[width=\columnwidth]{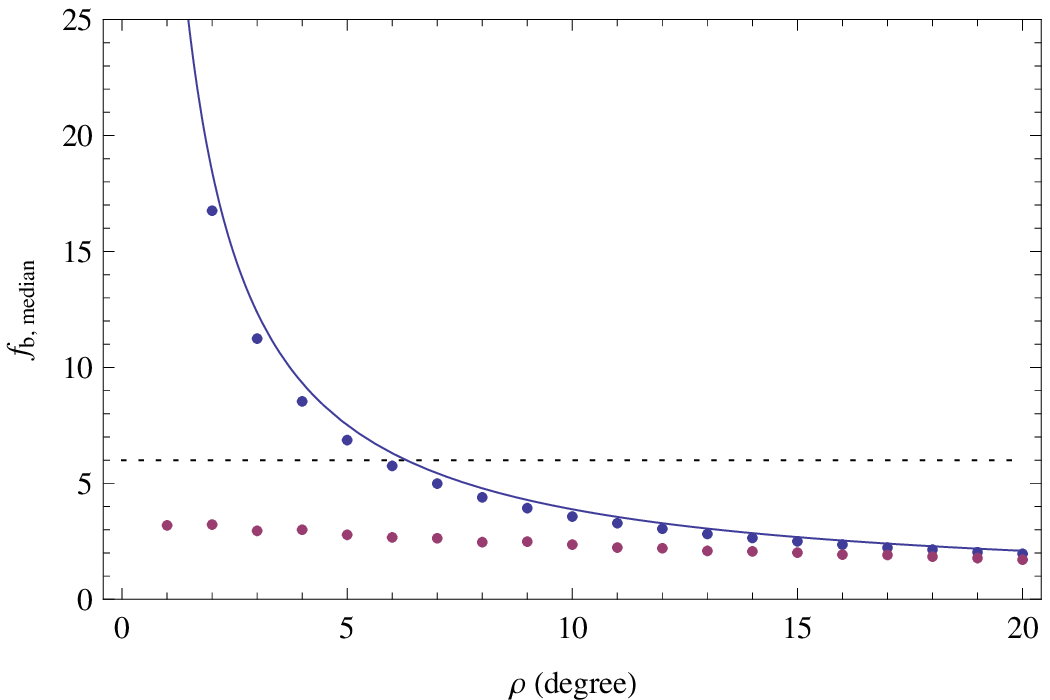}
\caption{\label{fig:Known:FbVersusRho} The median value of $f_{\rm b}$ for detectable pulsars (red; from $\pdf_e$  in Eq.\ (\ref{eq:pdf:Observed})) and all pulsars
  (blue; from $\pdf$  in Eq.\ (\ref{eq:pdf:Intrinsic})) versus $\rho$, assuming  a synthetic pulsar population with misalignment angle $\alpha$  uniformly
  distributed and beaming described by Eq. (\ref{eq:def:fb}).  
Also shown is $1/f_{\rm b} = (1-\cos\rho) + (\pi/2 -\rho)\sin\rho$, an estimate of $f_{\rm b}(\rho)$ assuming $\alpha$ is random; see  Emmerling and Chevalier (1989). 
The \emph{detectable} pulsar population will preferentially have wide beams.   Only a population of pulsars with
  very tight opening angles are consistent with the canonical value of $f_{\rm b}=6$.
 In our ``standard'' model (Eqs.\ 7 and 8), such
  narrow beams are improbable for pulsars
  spinning with   $10 \unit{ms}<P_s<100\unit{ms}$ that is relevant to most of the known PSR-NS binaries (see Table \ref{tab:psrns}).
 }
}
\end{figure}

\subsection{Partial information and competing proposals }
In a few cases, observations ${\cal O}_i$ of pulsar $i$  provide some information about that pulsar's geometry, such as
confidence intervals on $\alpha$ and $\rho$ or even a posterior distribution function $\pdf(\alpha,\rho|{\cal O}_i)$.
When a unique, superior constraint exists, we could have simply factored this information into the average that defines
$F=1/\fbeff$ (Eq.\ 9)  (e.g., $\alpha$ constraints for 
PSR J1141-6545; the preferred geometries of PSRs B1913+16 and B1534+12).
However, in cases where competing proposals exist (e.g., the two models for PSR J0737-3039A suggested by
\citet{0737Aspin1} and \citet{0737Aspin2}, also reviewed in \citet{doublepsrreview08}; see \S \ref{sec:Rates:PSRNS:Merging}), or where our prediction is
extremely sensitive to unavailable priors, such as the mean misalignment angle  of PSR J1141-6545,
\citep{J1141alphapaper,J1141alphakramer} (\S \ref{sec:sub:Rates:PSRWD}), we explicitly provide multiple solutions in the
text.   When multiple competing predictions are available, our final prediction averages over a range of predictions between them.  
 Explicitly, if $\pdf_f(\log f_{\rm b})$ is a  distribution in $X\equiv \log (f_{\rm b}/f_{\rm b,eff})$ reflecting our \emph{a priori} model uncertainty in $\fbeff$, and $\pdf(\log {\cal R}|f_{\rm b})= {\cal R} \pdf({\cal R}|f_{\rm b}) \ln 10$ is our posterior   estimate for the birthrate ${\cal R}$ given known beaming $f_{\rm b}$ [cf. Eq. \ref{eq:Posterior:Rate:KnownGeometry}], then a posterior estimate that reflects uncertain beaming is
\begin{eqnarray}
\label{eq:marginalization}
\pdf(\log {\cal R})d \log {\cal R}&=&\int dX \pdf_f(X+\log \fbeff) \pdf(\log {\cal R} - X |\fbeff)~.
\end{eqnarray} 
%

For the many PSR-NS binaries which have spin period between 10 ms and 100 ms,we calculate P(log R) using Eq. (11). Specifically, we average P(log R) over the beaming correction factor between $f_{\rm b,eff}$ (listed in  Table \ref{tab:psrns}) and the canonical value of 6, adopting a systematic error distribution $P_{sys}$(log $f_{\rm b}$) which is uniform in log $f_{\rm b}$ between these limits. For pulsars with spin period outside of this range, we use Eq.\ (1) with $f_{\rm b,eff}$ listed in Table \ref{tab:psrns} or \ref{tab:psrwd}.


\optional{
\subsection{Model errors (*)}

** key point to emphasize: tight beams can be possible, but so long as they are RARE they will never be seen and not
impact our prediction for the inferred number.

** this has impact on modeling errors: we can fail to model the probability of tight-beam systems...but that won't matter.
}


\input{tab-manual-KnownPulsars-PSRNS-deluxetable.tex}  

\begin{figure}
\includegraphics[width=\columnwidth]{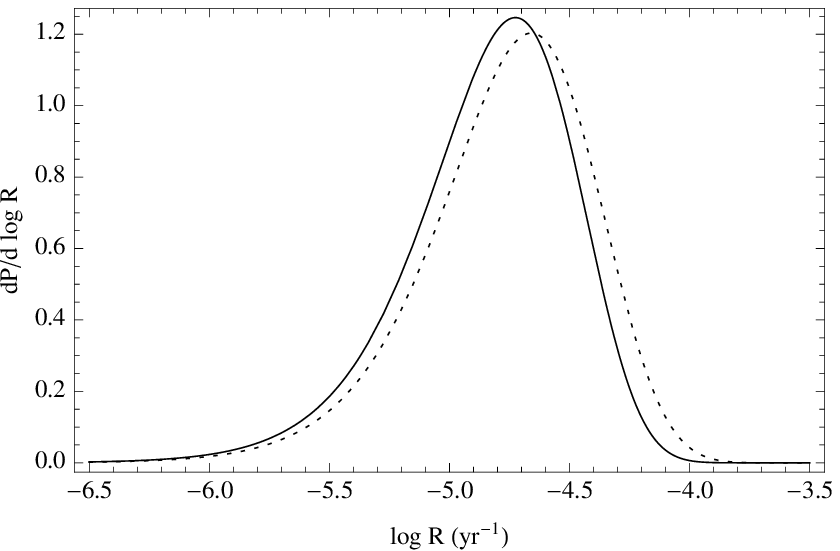}
\caption{\label{fig:Rates:PSRNS:Merging:Special0737WithLifetimeUncertainty}  Solid curve: our estimate for the birthrate
  of pulsars like PSR J0737-3039A, based on our standard beaming geometry model ($\fbeff=1.55$) and its spindown age
  ($\tau_{age}=145\unit{Myr}$).  Dotted curve: as previously, but marginalized over uncertainties in the age
  $\tau_{age}$.  Motivated by averaging together models 1,4, and 5 in \cite{J0737age}, assuming each model equally likely,  we approximate
  age uncertainty in PSR J0737-3039A by a uniform distribution in $t$ between 50 and 180 $\unit{Myr}$.  
While significantly shorter lifetimes are permitted, the median birthrate for pulsars like PSR J0737-3039A increases by only 
$20\%$,  from 18\permyr to  22 \permyr.
}\end{figure}

\begin{figure}
\chooseBorC{
\includegraphics[width=\columnwidth]{fig-mma-paper-Rates-PSRNS-Merging-BW}
}{
\includegraphics[width=\columnwidth]{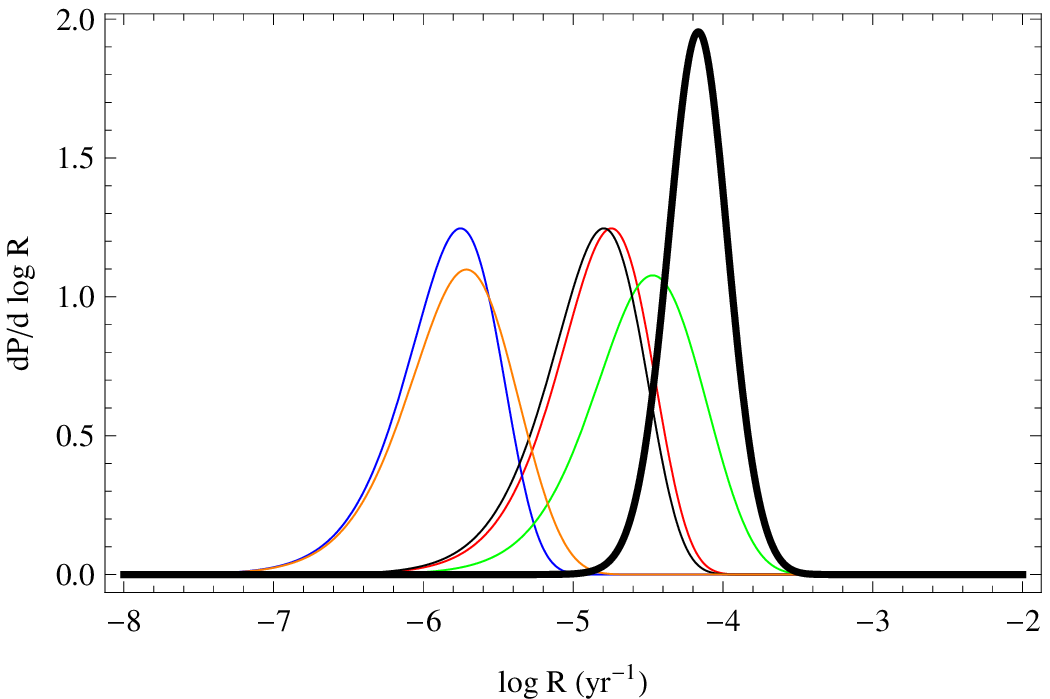}
}
 \caption{\label{fig:Rates:PSRNS:Merging} 
 Distributions $\pdf(\log {\cal R})$  for the birthrate of tight PSR-NS binaries: PSRs B1913+16\chooseBorC{ (dashed center)}{ (red)},
 B1534+12\chooseBorC{ (dashed left)}{ (blue)}, J0737-3039A\chooseBorC{ (dotted right)}{ (green)}, J1756-2251\chooseBorC{ (dotted left)}{ (orange)}, and J1906+0746\chooseBorC{ (dotted center)}{ (black)}.  The thick  black curve indicates our best estimate for the overall
 Galactic birthrate  of tight PSR-NS binaries, assuming the reference pulsar population model in \KimProceedings{}; the
 median value is $\simeq 89~\unit{Myr}^{-1}$. The individual 
\chooseBorC{}{colored} curves indicate our best estimates for the birthrate
 of the individual binaries, based on Eq.\ (1) and incorporating $\fbeff$  listed in  Table \ref{tab:psrns}. 
For the two PSRs with $10~\unit{ms} < P_{\rm s} <100~\unit{ms}$ and without specified beam geometry (PSRs J0737A-3039 and J1756-2251), we account for uncertainty in opening-angle modeling by allowing $\ln f_{\rm b}$ to be anywhere between
our `conservative' and `standard' choices. 
 }
%
%
%
\end{figure}


\input{tab-manual-KnownPulsars-PSRWD-deluxetable.tex} 

%
Similarly, if the lifetime is uncertain, one can marginalize $P(\log {\cal R})$ over the lifetime $\tau$; if the relative likelihood of different lifetimes $\pdf_{\tau}(\log \tau_i)$ is known, then defining $Y=\log ({\tau}/\tau_{i})$,
\begin{eqnarray}
\label{eq:marginalization_over_age}
\pdf(\log {\cal R})d \log {\cal R}&=&\int dY \pdf_\tau(Y + \log \tau_i) \pdf(\log {\cal R} + Y |\tau_i )~,
\end{eqnarray}
 Though usually the lifetime is relatively well determined, being dominated by relatively well determined merger or death timescale, we do use this expression to marginalize over the considerable uncertainty in the current age of PSR
  J0737-3039A, based on the proposed range of lifetimes presented by  \cite{J0737age}; see Fig.\ \ref{fig:Rates:PSRNS:Merging:Special0737WithLifetimeUncertainty}.

\section{RESULTS}
\subsection{\label{sec:Rates:PSRNS:Merging}Tight PSR-NS binaries}
The pulsar binaries we consider in this work and the estimated $\fbeff$  based on our standard model are summarized in  Table \ref{tab:psrns} and \ref{tab:psrwd}. 
Tight PSR-NS binaries contribute to our estimate of the overall PSR-NS birthrate (which, for these short-lived systems, is equivalent to their merger rate).
%
Among this set, both PSRs B1913+16 and B1534+12 have both $\rho$ and $\alpha$ measurements.  For PSR B1913+16, we adopt
$f_{b,obs}=5.72$ based on $\rho=12.4^{\circ}$\citep{WeisbergTaylor02}\footnote{\citet{krameretal98} also obtained
  similar value for $\rho$} and $\alpha=156^{o}$.   For PSR B1534+12, we use $f_{b,obs}=6.04$ based on
$\rho=4.87^{\circ}$ and $\alpha=114^{o}$ \citep{Arzoumanian1996B1534}. [The canonical value of $f_{\rm b}\sim6$ is
obtained from the average of the $f_{b,obs}$ for these pulsars.]   
Alternative choices for $\alpha$ and $\rho$ have been
proposed for both pulsars (Fig.\ \ref{fig:fig-mma-Models-fbEffVersusP}); combining the most extreme of these options
leads to values comparable to  our model's preferred values: $\fbeff=$ 2.2 and 1.8 for PSRs B1913+16 and B1534+12
respectively.   
As tightly beamed pulsars are unlikely to be seen in our fiducial or tightly beamed  model
[the solid and dashed curves in Fig. \ref{fig:fig-mma-Models-fbEffVersusP}],\footnote{As demonstrated here with $\rho$ and in the conclusions with
  $\alpha$ [Table \ref{tab:Conclusions}], the distribution of $\pdf(\rho,\alpha)$ must change dramatically to lead to a significant probability of
  detecting a pulsar with $f_b=6$.  Rather than introduce strong assumptions and large systematic errors to enforce it,
  biasing our expectations about pulsars unlike PSRs B1913+16 and B1534+12 but similar to well-constrained isolated
  pulsars, we instead adopt a parallel approach.  Our final results average between empirically-motivated theoretical
  priors, valid for all spin periods, and the assumption $f_b=6$, applied to pulsars similar to PSRs B1913+16 and
  B1534+12, with $10 \unit{ms}<P_s<100\unit{ms}$.
}
  we adopt an alternative approach for pulsars
with spin periods in the otherwise poorly constrained region  $10$ and $100\unit{ms}$, the interval containing most
PSR-NS binaries.

Pulsars in tight PSR-NS binaries may have wide opening angles, given their spin periods (see Fig.\
\ref{fig:Ingredients:SinglePulsarOpeningAngles}). If so, these binaries likely have  $f_{\rm b}<6$. However,  the
observations suggest that PSRs B1913+16 and B1534+12 have narrower beams (see Fig.\
\ref{fig:Ingredients:SinglePulsarOpeningAngles}, region defined by dotted lines). If these  narrower opening angles are
characteristic of tight, recycled pulsars in PSR-NS binaries, the appropriate  $f_{\rm b}$ could be closer to $6$.  In
particular, given the similarity in spin period of  PSR J0737-3039A to those pulsars and the lack of other constraints
in that period interval (see Fig.\ \ref{fig:Ingredients:SinglePulsarOpeningAngles}), we assume $\log f_{\rm b}$ could
take on any value between $\log 1.5$ (the value we estimate in our spin model) and $\log 6$.   
%


\begin{figure}
 \includegraphics[width=\columnwidth]{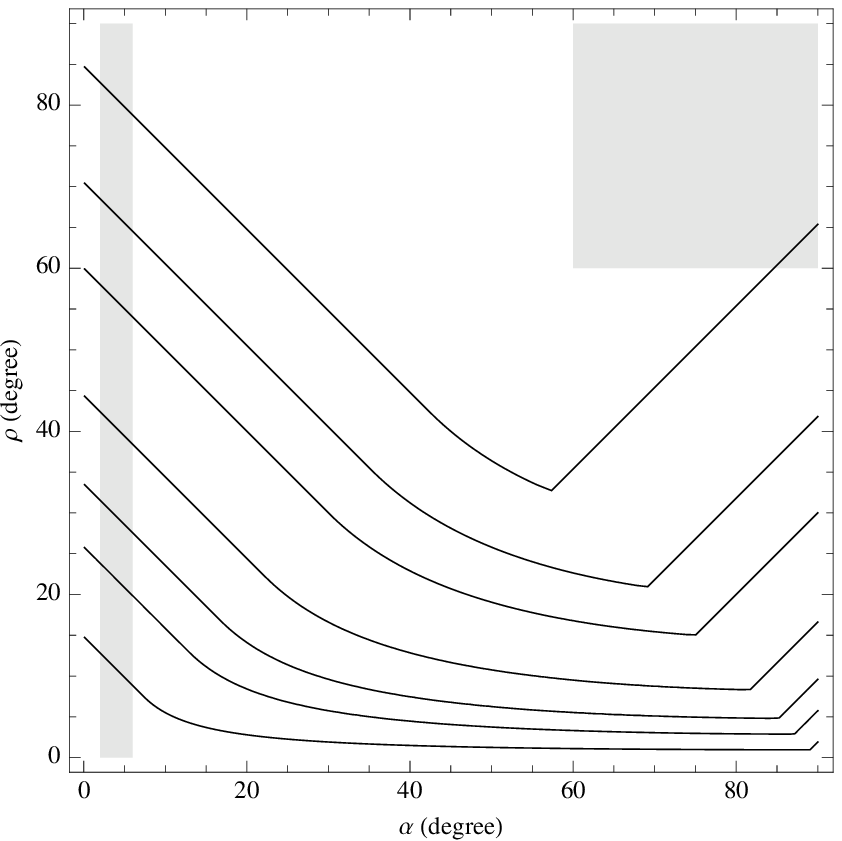}
 \caption{\label{fig:fig-mma-ExtremeBeaming-Contours} Plausible beaming in PSR J0737-3039A: Solid curves are contours $f_{\rm b}(\alpha,\rho)$ corresponding to
  $f_{\rm b}=1.1,1.5,2,3.5, 6,10,30$ (top to bottom). 
 Overlaid are  approximately-drawn regions of  $\alpha$ and $\rho$ which are consistent with  observations of PSR J0737-3039A,  corresponding to the unipolar model (left box, small $\alpha$)
 and two-pole model  (right box);  compare with contours and discussion in
 \citet{0737Aspin1}. 
 }
\end{figure}

Given  posterior likelihoods, we could explicitly and systematically include  observational constraints on the beaming geometry of PSR J0737-3039A as described earlier; see, e.g., the posterior constraints in \citet{0737Aspin1} and \citet{0737Aspin2}.  Observations support two alternate scenarios.  In one, the pulse is interpreted as from a single highly aligned pole ($\alpha < 4^\circ$).  Because of its  tight alignment, in this model the beaming correction factor should be large: at least as large as those for binary pulsars ($f_b\simeq 6$, assuming $\rho \sim 30^\circ$, from $\rho(P_{\rm s})$), and potentially  larger ($f_b\simeq 30$ assuming $\rho =10^\circ$, based on observed opening angles for PSRs B1913+16 and B1534+12).  In the other scenario, favored by recent observations \citep{0737Aspin2}, the pulse profile is interpreted as a double pole orthogonal rotator $\alpha \simeq \pi/2$ with a fairly wide beam ($\rho\sim$60$^{\circ}-$90$^{\circ}$, consistent with $\rho(P_{\rm s})$).  This latter case is  consistent with our canonical model and leads to a comparable $f_b$.  Comparing with the assumptions presented  earlier, so long as we ignore the possibility of tight alignment and narrow beams,  our prefered model and uncertainties for PSR J0737-3039A already roughly incorporate its  most significant modeling uncertainties.
%
 Considering that the contribution from PSR J1906+0746 is comparable with that of the PSR J0737-3039A, our best estimate for the birthrate of merging PSR-NS binaries is not very sensitive to changes in a nearly orthogonal-rotator geometry model for PSR J0737-3039A.
However, because we cannot rule out the most extreme scenarios for PSR J0737-3039A, for completeness we also describe 
implications of a unipolar model: 
the beam shape constraints summarized by Fig.\
\ref{fig:fig-mma-ExtremeBeaming-Contours} translate to a prior on 
$\log f_b$ that is roughly uniform between $\log 6$ and $\log 30$.

 In Fig.\ \ref{fig:Rates:PSRNS:Merging}, we show that the probability distribution of PSR-NS merger rate with our best
 estimates for the beaming correction $f_{\rm b,eff}$, assuming the reference model of \KimProceedings{}.   $P({\cal
   R}$)'s follow from convolving together birthrate distributions based on each individual pulsar binary,  where those
 birthrate distributions are
 calculated as described in previous sections.   Including PSR J1906+0746, we found the  median PSR-NS merger rate is
 $\simeq 89$ Myr$^{-1}$, which is smaller than what we predicted in \KimProceedings{} ( $\sim123$ Myr$^{-1}$, 
   cf.\ their peak value is 118 Myr$^{-1}$), assuming the same $\tau_{age}$ and $\tau_{mrg}$  listed in Table \ref{tab:psrwd}, but used
 $f_{b,J0737}=5.9$,  
due entirely to the smaller beaming correction factors for PSR J0737-3039A allowed for in this work.    Though our best
estimate for the merger rate is slightly smaller than previous analyses, the difference is comparable to the
Poisson-limited birthrate uncertainty and much smaller than the luminosity model uncertainty described in the appendix.
Being nearly unchanged, our study has astrophysical implications in agreement with prior work such as \PSmoreconstraints and
\cite{PSellipticals}.

Although the merging timescale is reletively well-defined, we note that our estimate does not include $O(30\%)$
uncertainties in the current binary age.  In this work, for example, we fix the total age of PSR J0737-3039A to be $\sim$230 Myr. 
 Fig.\  \ref{fig:Rates:PSRNS:Merging:Special0737WithLifetimeUncertainty} shows a marginalized P(${\cal R}$) using
  the age constraints from \citep{J0737age} between 
50 and 180 $\unit{Myr}$   (These models take into account interaction between the
  two pulsars; we omit the two most extreme models 2 and 3 with rapid magnetic field decay).

%


\subsection{Wide PSR-NS binaries}
\label{sec:sub:WidePSRNS}
In addition to those in tight orbits, some PSR-NS binaries have wide orbits (orbital period is larger than 10
hours). These wide binaries would never merge through gravitational radiation within a Hubble time.  Their estimated death timescales imply that most of the known pulsars in wide binaries will remain visible for time comparable to or in excess of $10~\unit{Gyr}$ (see  Table \ref{tab:psrns}) and  that binaries have accumulated over the Milky Way to their present
number.  In this work, we assume that PSR J1753-2240 \citep{J1753discovery} is another wide PSR-NS
  binary.  Motivated by the fact that spin periods of pulsars found in wide orbits are comparable to those relevant to
tight PSR-NS binaries, we adopt the same techniques for estimating $\fbeff$. For PSRs J1811-1736 and J1753-2240, we
adopt the  $\fbeff$ values expected from their spin periods. On the other hand, for PSRs J1518+4904 and J1829+2456, we
average $\pdf({\cal R}|f_{\rm b})$ over a range of $\ln f_{\rm b}$  between the value predicted ($\ln \fbeff$ 
  listed in Table \ref{tab:psrns}) and $\ln 6$.  Our estimate for the birthrate of ``wide'' PSR-NS binaries, shown in Fig.\
\ref{fig:Rates:PSRNS:Wide}, is slightly higher than previously published estimates by
\PSmoreconstraints{}. 
The birthrate of wide PSR-NS binaries is not significantly changed due to the discovery of PSR J1753-2240, because of its resemblance to PSR J1811-1736. Contrary to \S~\ref{sec:rateconst}, the previous analyses had assumed that the pulsar population reached number equilibrium, e.g., \citet{PSconstraints1}; our effective lifetime is 2 times shorter.  At the same time, our beaming correction is roughly three times smaller, so the
relevant factors mostly cancel; our median prediction is almost exactly 1.5 times lower than the previous estimate.  


The reconstructed $P({\cal R}$) assuming steady-state star formation  has a median at 0.84 Myr$^{-1}$ and is narrower than the previous work due to the discovery of PSR J1753-2240, with a ratio between the upper and lower rate estimate at 95\% confidence interval, ${\cal R}_{95\%}/{\cal R}_{5\%}$, is 8.8 (compared to 13.8 without J1753-2240).
As in \PSmoreconstraints, the birthrate estimate of ``tight'' PSR-NS binaries is  still roughly 100 times greater than
that of ``wide'' binaries, even though both estimates have been modified through lifetime and $\fbeff$ corrections. Interestingly, known ``wide'' PSR-NS binaries were discovered through recycled pulsars exclusively. 
Following \PSmoreconstraints, we interpret the difference in birthrate as evidence for a  selection bias: only a few progenitors of  PSR-NS binaries undergo enough mass transfer to spin up the PSR, but not enough to bring the orbit close enough to merge.

\begin{figure}
\chooseBorC{
 \includegraphics[width=\columnwidth]{fig-mma-paper-Rates-PSRNS-Wide-BW}
\caption{\label{fig:Rates:PSRNS:Wide}
 Same as Fig.\ \ref{fig:Rates:PSRNS:Merging}, but for the wide PSR-NS population: PSRs J1811-1736 (dotted center),  J1518+4904
 (dotted left), J1829+2456 (dashed left), and J1753-2240 (dashed center), along with our best guess for the overall rate (thick solid
 curve).  The median birthrate for wide PSR-NS binaries is $\simeq 0.84~\unit{Myr}^{-1}$, assuming the preferred pulsar
 population model in \KimProceedings{} and a steady star formation rate.  The newly discovered PSR J1753-2240 is a cousin of PSR J1811-1736:  their contributions to the Galactic birthrate are comparable.   
 }
}{
 \includegraphics[width=\columnwidth]{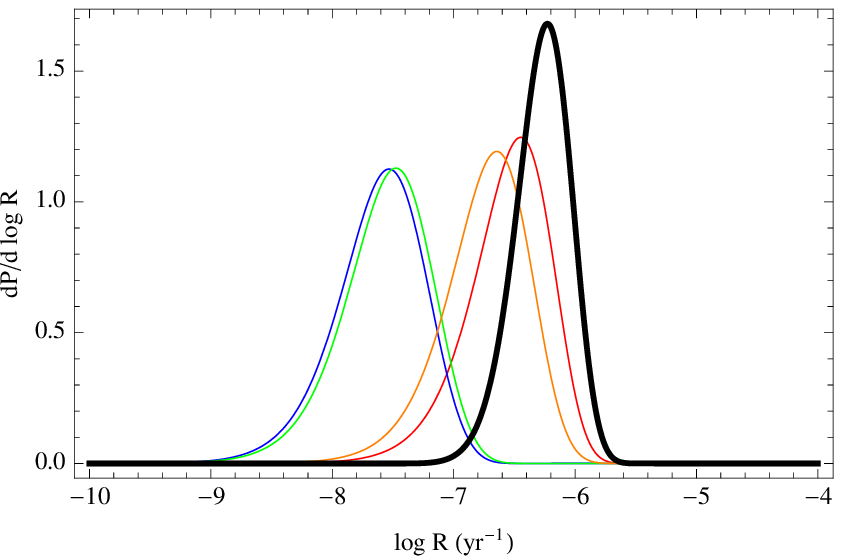}
\caption{\label{fig:Rates:PSRNS:Wide}
 Same as Fig.\ \ref{fig:Rates:PSRNS:Merging}, but for the wide PSR-NS population: PSRs J1811-1736 (red),  J1518+4904
 (blue),  J1829+2456 (green), and J1753-2240 (orange), along with our best guess for the overall rate (thick black
 curve).  The median birthrate for wide PSR-NS binaries is $ 0.84~\unit{Myr}^{-1}$, assuming the preferred pulsar
 population model in \KimProceedings{}  and a steady star formation rate.  The newly discovered PSR J1753-2240 is a cousin of PSR J1811-1736:  their contributions to the Galactic birthrate are comparable. 
}
}
\end{figure}

\subsection{\label{sec:sub:Rates:PSRWD}Tight PSR-WD binaries}
\begin{figure}
 \includegraphics[width=\columnwidth]{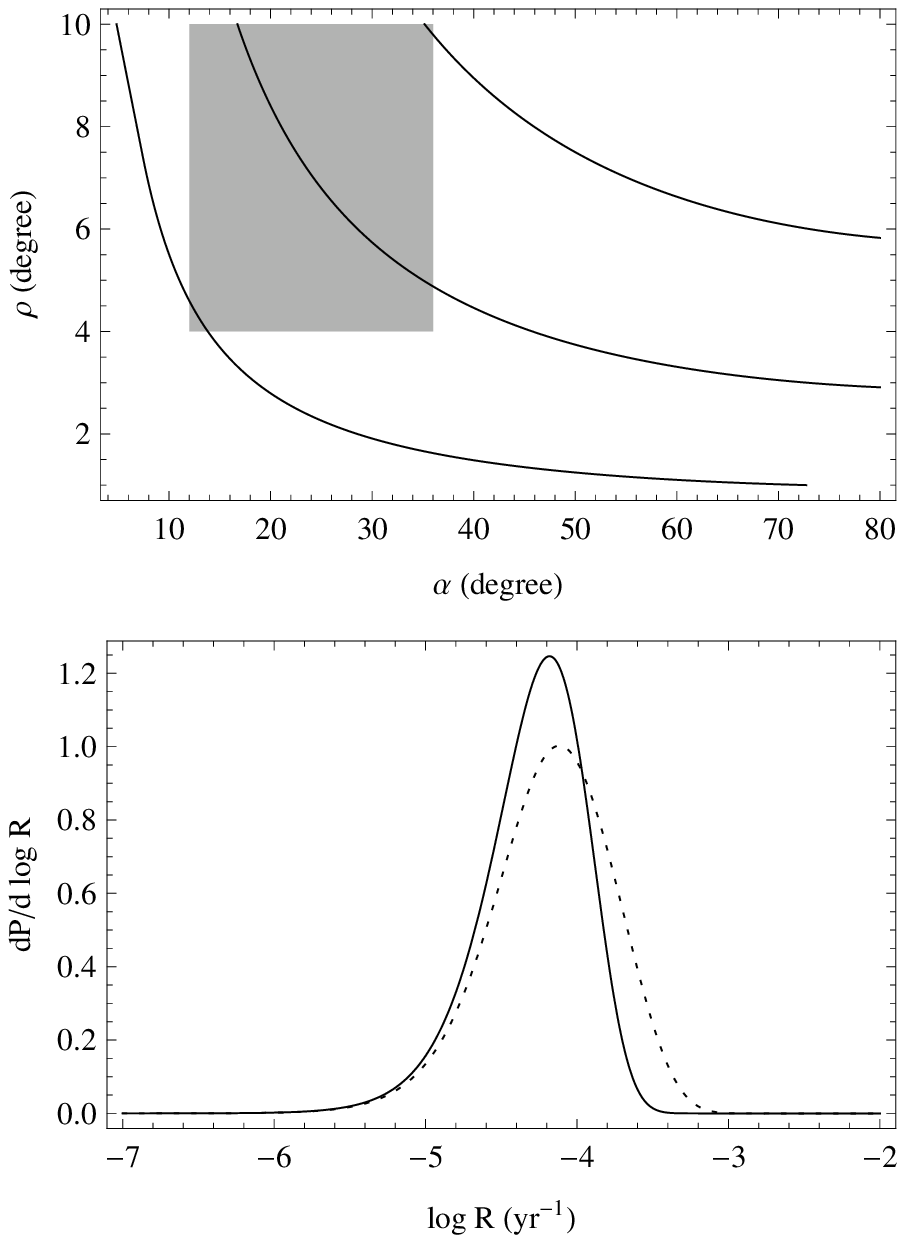}
\caption{\label{fig:Rates:PSRWD:1141Only} Top panel:  A contour plot of $f_{b}(\alpha,\rho)$ over the range relevant to PSR J1141-6545; contours shown are $f_{b}=5,10,30$ from top to bottom, respectively. The shaded region indicates the range of $\alpha$ and $\rho$ implied by a preliminary analysis of recent observations \cite{J1141alphapaper,J1141alphakramer}; inside this box $f_{\rm b}$ ranges between 5 and 30.  Bottom panel: $\pdf({\cal R})$ for PSR J1141-6545 based on our standard model ($\fbeff=$ 5.46; solid,  also see Table \ref{tab:psrwd}) and the constraints above (averaging $\fbeff$ between 5.46 and 30 using Eq.\ (\ref{eq:marginalization}); dashed). 
}
\end{figure}

With the exception of PSR J1141-6545, the lifetimes of tight PSR-WD binaries are limited by their gravitational wave inspiral time \citep{peters64}. The net visible lifetimes of these binaries are  significantly in excess of the age of the Milky Way, and are strongly recycled with $P<10~\unit{ms}$.   Based on comparison with measured isolated pulsars, a likely range for their opening angles and thus $\fbeff$ can be estimated  (see Table \ref{tab:psrwd} for a summary).   Combining this information, birthrates for the three millisecond PSR-WD binaries, PSRs J0751+1807, J1757-5322, and the newly discovered J1738+0333, are easily estimated (Fig.\ \ref{fig:Rates:PSRWD}).
The birthrate for tight PSR-WD binaries, however, is dominated by the non-recycled pulsar PSR J1141-6545.  For this unusual binary, 
 the visible pulsar lifetime is dictated by its death timescale ($\tau_{\rm d}$), rather than its gravitational-wave merger time ($\tau_{\rm mrg}$), e.g., \citet{kklnswd}.  
%
Based on different models for the death line, we expect the  total lifetime of PSR J1141-6545 to be between between $100
- 170$ Myr;  
 for the purposes of a rate estimate, we allow $\tau$ to be logarithmically distributed within these limits.
 Additionally,  recent observations imply its misalignment angle to be   $160^{\circ}$$^{+8}_{-16}$ (or equivalently,
 $12^\circ<\alpha < 36^\circ$, which is adapted in this work) at a 68\% confidence level,
 \citep{J1141alphapaper,J1141alphakramer}; see Fig.\ \ref{fig:Rates:PSRWD:1141Only}.  Without this alignment information
 we would already predict fairly substantial  beaming: $\fbeff\approx 5.46$   solely based on $\rho(P_{\rm s})\approx 8{^\circ}.6$  (Table \ref{tab:psrwd}).  With a nearly-polar beam, however, the beaming could be $10$ times tighter.  We therefore present two estimates for the birthrate of PSR J1141-6545, contingent on $\alpha$: our current \emph{a priori} model, shown as a solid curve in Fig.\ \ref{fig:Rates:PSRWD:1141Only}; and an estimate that includes tight beaming (dotted curve).


\begin{figure}
\chooseBorC{
 \includegraphics[width=\columnwidth]{fig-mma-paper-Rates-PSRWD-Merging-BW}
 \caption{\label{fig:Rates:PSRWD}
 Empirical distributions $\pdf(\log {\cal R})$ implied by the set of known, tight PSR-WD binaries:  PSRs J1141-6545
 (dotted right),
 J1757-5322 (dashed left), J1738+0333 (dotted left, superposed), and J0751+1807 (dashed center) given the assumptions about opening angle and
 misalignment. The Galactic P(${\cal R}$) is also shown in a thick solid curve. The P(${\cal R}$) for PSR J1141-6545
 presented here is obtained by our standard prediction: assuming the pulsar is one of a family of pulsars with
 misalignment angles $\alpha$ uniformly distributed between $0$ and $\pi/2$ and $\rho$ normally distributed as in Fig.\
 \ref{fig:Ingredients:SinglePulsarOpeningAngles} (cf.\ see Fig.\ \ref{fig:Rates:PSRWD:1141Only}).  
 Based on our standard beaming geometry model, PSR J1141-6545 still dominates the Galactic birthrate of PSR-WD binaries. 
The median rate for the Galactic PSR-WD binaries (thick black solid curve) is estimated to be $\simeq 34~\unit{Myr}^{-1}$. 
 }
}{
 \includegraphics[width=\columnwidth]{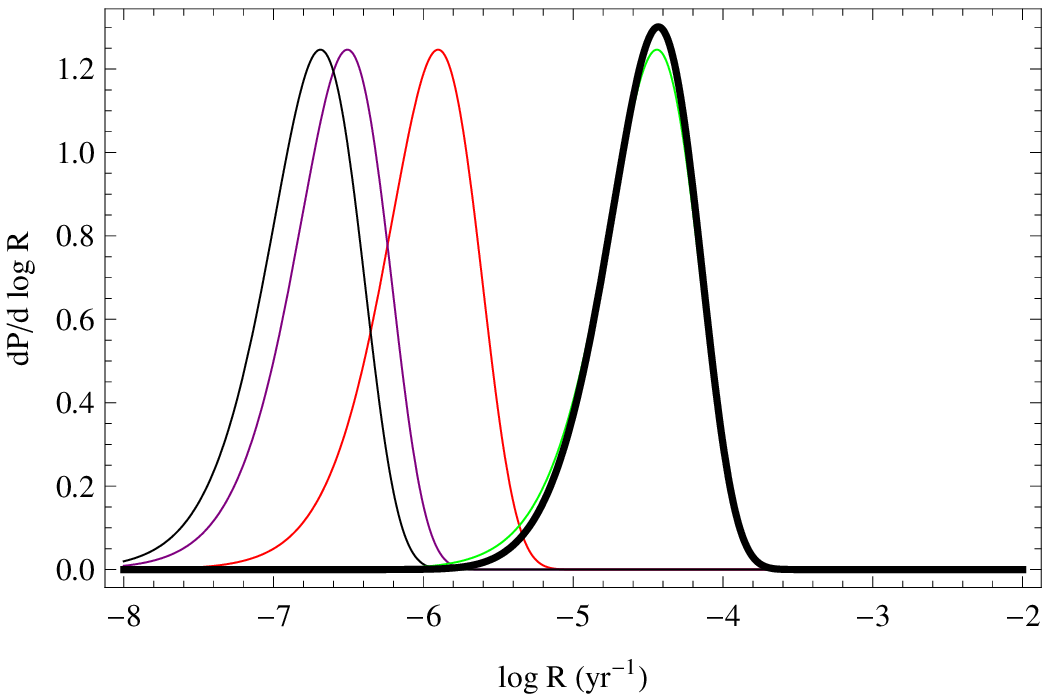}
\caption{\label{fig:Rates:PSRWD}
 Empirical distributions $\pdf(\log {\cal R})$ implied by the set of known, tight PSR-WD binaries:  PSRs J1141-6545
 (green), J1757-5322 (purple), J0751+1807 (red), and J1738+0333 (blue), given the assumptions about opening angle and
 misalignment.
The P(${\cal R}$) for PSR J1141-6545 presented here (solid green curve) is obtained by our standard prediction: assuming
the pulsar is one of a family of pulsars with  misalignment angles $\alpha$ uniformly distributed between $0$ and $\pi$
and $\rho$ normally distributed as in Fig.\ \ref{fig:Ingredients:SinglePulsarOpeningAngles} (cf.\ see Fig.\
\ref{fig:Rates:PSRWD:1141Only}).   Based on our standard beaming geometry model, PSR J1141-6545 still dominates the
Galactic birthrate of PSR-WD binaries.  The median rate for the Galactic PSR-WD binaries (thick black solid curve) is
estimated to be $\simeq 34~\unit{Myr}^{-1}$
 }
}
\end{figure}

 Pulse profiles are an essential ingredient in empirical pulsar population modeling. 
The pulse profile of PSR J1141-6545 has  significantly broadened, by a factor 8, since the pulsar was discovered in 2000 \citep{J1141alphakramer}. 
This is mainly attributed to the effects of the geodetic precession. Recalculating  $N_{psr}$ for PSR J1141-6545
  with the current, broader pulse, the most likely value of  $N_{psr}$ is now $\sim$1000 instead of 350 that is
  estimated with the pulse profile in 2000 (see \citet{kklnswd} for the details).\footnote{This preliminary result does
  not include up-to-date
Doppler smearing, consistent with the wider pulse width.}
We expect that an analysis that treats both geodetic precession and the duration of pulsar radio surveys imply a
   slightly different birthrate for the PSR J1141-6545 
than what has been shown in \citet{kklnswd}.
 In this work, however, we use the pulse profile presented in the discovery paper \citep{J1141discovery} and assume no evolution in the pulse probile, as we only focus on the effects of the beaming
correction factor with various beam geometry models. The effects of the pulse profile evolution of PSR J1141-6545 to the Galactic birthrate of PSR-WD binaries, taking into account more detailed corrections for observational biases such as the Doppler smearing, will be discussed in a separate paper (Kim et al.\ in prep).  



Our preferred birthrate estimate agrees with previously published estimates based on a ``fidicual'' beaming factor
$\fb=6$ [\PSmoreconstraints; compare $C=53\unit{kyr}$ here with $A=47\unit{kyr}$ in their Table 1 for PSR J1141-6545],
as well as with early estimates that adopt $\fb=1$ [\cite{kklnswd}  predict a \emph{peak-probability} galactic birthrate of $4
  f_b$\permyr, based on an $N_{psr}=400$].\footnote{When describing preferred birthrates, we cite  \emph{median}
    probability. Earlier  papers like
    \KimKalogeraLorimer{} and \cite{kklnswd}  instead cite \emph{peak of the PDF} $p(R)$.  For the lognormal and poisson
    distributions typical in this problem, the two disagree: the median is generally slightly higher.}   By coincidence
the fiducial beaming factor agrees with our best beaming estimates based on $\rho(P)$ data for comparable-spin pulsars.
Our birthrate estimate also agrees with a recently published independent estimate of PSR-NS (or PSR-BH) birthrates based
on optical discovery of the tight WD-compact object binary SESS 1257+5428  described in \cite{obs-bin-wd-EmpiricalHighMergerRate2009}.

\subsection{Discussion}

\input{tab-mma-KnownPulsars-MisalignmentOptions-CKrevised.tex}

\noindent \emph{Alternate misalignment models}:
In order to calculate $\fbeff$, we adopt a flat distribution as our standard model for the $\alpha$ distribution defined in Eq.\ (\ref{eq:alphaflat}). 
Several alternative distributions for pulsar misalignment have been proposed; some, however, do not self-consistently
account for detection bias.  For example, a randomly aligned beam vector on the sphere ($p(\alpha)d\alpha=d\cos\alpha$)
implies that nearly all  \emph{ detected} pulsars should be nearly orthogonal rotators. 
The  observation of many pulsars with smaller $\alpha$, e.g., \citet{kol04}, is not consistent with that distribution.   On the other hand,  a flat
 distribution and even more centrally concentrated ones (e.g., $p(\alpha)=\cos\alpha d\alpha$) are plausible, considering observational biases can explain the lack of detected pulsars with small
 $\alpha$ (say, $<$ 45$^{o}$ from Fig.\ $3$ in \citet{kol04}).   But because the expectation defining  $\fbeff$ (Eq.\ 9) is governed by the smallest typical beaming fraction of {\it detected}
 pulsars, the predictions for $\fbeff$ are fairly independent of misalignment model, assuming it is not too
 concentrated near the equator or poles.  For this reason, and without more information to quantify uncertainties in a
 reconstructed $\alpha$ distribution with which to base a comparison (e.g., 
the selection bias of having an $\alpha$ measurement), we choose a flat distribution for simplicity. 

\optional{
\noindent \emph{Alternate opening angle models}:  As emphasized repeatedly in the text, fai $\rho(P)$
}

\noindent \emph{What beaming to use?}:  No one beaming correction factor applies for all circumstances: even for a
\emph{fixed} beam geometry distribution, different ``natural'' choices $\fb$ make sense for different questions [Figure
\ref{fig:Known:FbVersusRho}].  Beaming depends strongly on spin.   Finally, the canonical choice $\fb=6$ implicitly requires exceptionally tight
beaming, strong alignment, or both [Figure
\ref{fig:Known:FbVersusRho} and Table \ref{tab:Conclusions}].   If one number must be adopted, use the  spin-dependent
relation proposed by \citet{1998MNRAS.298..625T}  for pulsars with $P>100 \unit{ms}$ [Figure  \ref{fig:fig-mma-Models-fbEffVersusP}].

\optional{
\noindent \emph{Prior on reconstruction}

 For an observed PSR, when constructing posteriors, favor systems with less
  beaming strongly. \editremark{chunglee}: ask Dunc or MK if this kind of prior is built in to their estimate for
  $\alpha,\rho$ or not.
}

\optional{
\noindent \emph{Astrophysical impact of updated rates}

* most significant change is PSR-WD rate, which is higher; PSR-NS rate, slightly lower.  All these, combined with added
systematic uncertainty, make it \emph{easier} 
to explain results in terms of binary evolution models.

}

\section{Conclusions}
In this paper, we revise estimates for the birthrate of Galactic pulsar binaries, including new  binaries as well as updated
binary parameters and uncertainties.
%
We describe a new quantity,  the ``effective beaming correction factor'' $\fbeff$, as a tool to permit reconstruction of
Galactic binary pulsar birthrates from observations, when observations support not just one choice but a
\emph{distributions} of pulsar beam geometries.    Currently, the best constrained $f_{\rm b}$'s are available
for PSRs B1913+16 and B1534+12, $f_{b,obs}\sim6$. 
Previous empirical birthrate estimates like \KimProceedings{} adopted this factor for all pulsar binaries, ignoring experience
from isolated pulsars' beams.
Instead, we adopt random misalignment angles and a fiducial choice for $\rho(P_{\rm s})$.  As summarized in Tables I and
II, this choice produces   $\fbeff$  significantly smaller than the canonical value of 6 for most pulsar binaries (cf.\ also see Table \ref{tab:Conclusions} for comparisons between differerent
beam geometry models), and for $P_{\rm s}>10\unit{ms}$ in agreement with the simple expression provided by \citet{1998MNRAS.298..625T}.
Generally, a detected pulsar should be a
priori  assumed to have nearly as large a beam (as small an $f_{\rm b}$) as its spin allows.  Although possible, tight beams cross our line of sight rarely.  Unless implicitly assumed ubiqutous, tightly beamed pulsars will not
significantly impact a birthrate estimate.
For pulsars with spin period between 10 ms and 100 ms, where few
observations exist to justify  extrapolations, we anchor our theoretical expectations for wide beams with observations
of comparable pulsar binaries that suggest strong beaming, averaging between the two extremes.  Most of the pulsar
binaries that dominate $P({\cal R})$ have spin periods in intervals where  $\rho(P_{\rm s})$ is well-sampled.   Of the
three classes of binaries considered, only the
birthrate of wide PSR-NS binaries is dominated by pulsars in the least-well-constrained interval
 $P_{\rm s}$ between $10$ and 100 ms.    Finally, the two binaries PSRs J1141-6545 and J0737-3039A
that 
dominate the birthrate of  tight PSR-WD and PSR-NS binaries, respectively, have constrained beam geometries.   In both
cases, the plausible beam geometries allow  a significantly narrower beam and thus a
correspondingly higher birthrate.   For these two pulsars and more generally for any pulsar binary 
with spin periods between $10$ and $100$ ms,  empirical posterior distributions 
$\pdf(\rho,\alpha)$, if made available, would help us better determine model-independent detected beam geometries and
therefore improve our estimates of pulsar binary birthrates.

Studies of nonrecycled pulsars (most with $P\simeq O(0.1-2~\unit{s})$) have suggested that isolated and binary pulsars
could align their spin and magnetic axes ($\alpha\rightarrow 0$) on $O(70, 200~\unit{Myr})$ timescales respectively
\citep{WJ08interpulses}. While spin-beam alignment leads to narrower beams and therefore significantly increased typical
$f_{\rm b}$ for older nonrecycled pulsars, almost all the pulsars studied here are recycled; the two nonrecycled PSRs
J1906+0746 and J1141-6545 are far too young for the proposed process to occur.   
At the other extreme, \cite{psr-ClaimRapidAlignment-Young2009} recently proposed extremely rapid alignment for
isolated pulsars, where  on short alignment timescales $\simeq 1 \unit{Myr}$ beams align and the  opening angles
converge to
$\rho_\infty \approx 2-5^o$.  If applied to the typical $\tau_i\simeq O(100\unit{Myr})$
binary pulsars considered here, this model implies  exceptionally strong beaming ($\fb \simeq 2/\rho_\infty^2 \simeq
O(200)$) and, for example, merging PSR-NS birthrates comparable to the galactic supernova rate.  
In a forthcoming paper we will address time-dependent effects like  alignment in pulsar binaries.
At present, our birthrate estimate relies only on observed pulsars as representatives of their evolutionary classes, without  any
correction for \emph{when} along their evolutionary track they have been detected (e.g.,  in
luminosity or, here, beam size; cf.\ \cite{1981MNRAS.194..137P}).

We find  similar binary pulsar birthrates as previous studies when we adopt the same reference pulsar population model as
\KimProceedings{}  and  our standard beaming geometry distribution.
For example, the median birthrate of tight PSR-NS binaries is $\simeq89$ Myr$^{-1}$;  
this estimate is lower than that of \KimProceedings{}, mainly due to the narrower beams required there.
Most pulsars in the
PSR-NS population have spin period between 10 ms and 100 ms, where few measurements of misalignment and opening angle
provide a sound basis for extrapolation.  For these pulsars, we anchor our prior assumptions on the ``fidicual'' value
$f_b\approx 6$ drawn from comparable-spin binary pulsars. 
We also updated the birthrate of wide PSR-NS binaries, including a newly
discovered binary, PSR J1753-2240.   Though we include improved estimates for the  effective lifetime and beaming for
each pulsar, these changes nearly cancel,
leaving our birthrate estimate similar to \PSmoreconstraints{}.   The discovery of PSR J1753-2240 still improves our understanding of P(${\cal R}$),
  adding redundancy and reducing the uncertainty in the birthrate.  For example, the ratio between the upper and lower
  limits of birthrate estimates at 95\% confidence interval is 8.8, including PSR J1753-2240: $1.5$ times smaller than
  the previous estimate ($13.8$) based on three binaries. 
Finally, we updated  the birthrate distribution of tight PSR-WD binaries, including the newly discovered PSR J1738+0333
\citep{J1738discovery}; PSR
J1141-6545 still dominates the total birthrate.    
Though PSR J1141-6545's pulse width  has evolved since its discovery, its \emph{discovery-time} pulse width best
characterizes how surveys found it.  Until a future attempt at time-dependent pulsar surveys, we adopt that reference
width when calculating its birthrate.
\optional{We also In order to calculate the {\it present} P(${\cal R}$) of PSR-WD
binaries, it is important to take into accout the effects of a time evolution of its pulse width due to the geodetic
precession (e.g. \citet{J1141alphapaper,J1141alphakramer}). If we consider the pulse broadening of PSR J1141-6545 since
its discovery in 2000 (Kaspi et al.\ 2000), we expect that PSR J1141-6545 would be even  more dominant.  
}

Two pulsars which dominate their respective birthrates have weakly constrained pulsar geometries: PSR J0737-3039A and
J1141-6545.  Both pulsars could be consistent with our standard beaming model; both could admit much narrower beams.
Information about these pulsars' geometries is \emph{not} included in our preferred birthrate estimates.  However,  in
the text we describe how our results change if updated beaming geometry information becomes available; see
Figures \ref{fig:fig-mma-ExtremeBeaming-Contours} and \ref{fig:Rates:PSRWD:1141Only}.

To facilitate comparison with previous results like \PSmoreconstraints{}, in the text we adopted a steady star formation
rate and did not marginalize over uncertainty in the parameters of our pulsar luminosity model.   Our best estimates 
ncluding these factors
are shown in Figure  \ref{fig:GlobalAmbiguity}; see the Appendix.
If future observations more tightly constrain the distribution $p(L_{min},p)$ of luminosity model parameters, these
final composite predictions can be easily re-evaluated using the information provided here.

\acknowledgments
CK is supported by a Marie-Curie International-Incoming Fellowship under the European Commission's FP7 framework. ROS is
supported by NSF award PHY 06-53462 and the Center for  Gravitational Wave Physics. The authors would like to thank
Michael Kramer for the misalignment measurements and pulse width evolution of PSR J1141-6545,
and Duncan Lorimer for his numerical data regarding the lifetime of PSR J0737-3039A, 
 his critical reading of the manuscript, and many useful suggestions. The authors also appreciate the hospitality of the Aspen Center for Physics, where this work commenced.


\bibliography{chunglee,tmprevised} 


\appendix

\section{Pulsar luminosity functions}
\subsection{Power law}
For clarity, in the text we discussed the impact of spin-dependent pulsar beaming on the birthrate given a fixed pulsar
population model. The results shown in this work are based on our reference model (a power-law distribution for a
pseudoluminosity $d\log N/dlog L =  -1$, with 0.3 mJy kpc$^{2}$ as the minimum intrinsic luminosity at 400MHz, Gaussian distribution in radial direction where the radial scale length is assumed to be $R_o=4$kpc, and exponential function in z, with a scale height of $Z_o=1.5$kpc).
See \KimKalogeraLorimer{}, \KimProceedings{} 
for further details on the pulsar population model. Although the rate estimates are sensitive to model parameters, or the fraction of faint pulsar assumed in a model, the qualitative feature described here are robust with different model assumptions made. As discussed in \KimProceedings{}, however, the reconstructed pulsar population depends sensitively on the luminosity function model. In this appendix, we quantify the uncertainties attributed to a pulsar luminosity function following \KimProceedings{}. For example, if the cumulative probability of a luminosity greater than $L$ is modeled by $P(>L) = (L_{min}/L)^{1-p}$, a function with two parameters $p$ and $L_{min}$, then empirically we have found the number of pulsar binaries $N_{psr}$ implied by  one detection scales as 
\begin{subequations}
\label{eq:AmbiguityInputs}
\begin{eqnarray}
N_{psr}&\propto& L_{min} 10^{-1.6 p} 
\end{eqnarray}
%
Assuming that a pulsar luminosity function is similar for both millisecond pulsars ($P_{\rm s} < 20$ms) and pulsars found in compact binaries, \KimProceedings{} adapt the results shown in \citep{cc97} and estimate that the uncertainty in $L_{min}$ and $p$ can be described by the two uncorrelated distributions
%
\begin{eqnarray}
\pdf(L_{min})&=& 1.22 L_{min}(1.7-L_{min}) \quad L_{min}\in [0, 1.7]\unit{mJy}\unit{kpc}^2\\
\pdf(p)&=& \frac{1}{\sqrt{2\pi\sigma_p^2}}e^{-(p-2)^2/2\sigma_p^2}   
   \quad \sigma_p =0.12  \quad  p \in [1.4, 2.6]
\end{eqnarray}
\end{subequations}

\begin{figure}
\includegraphics{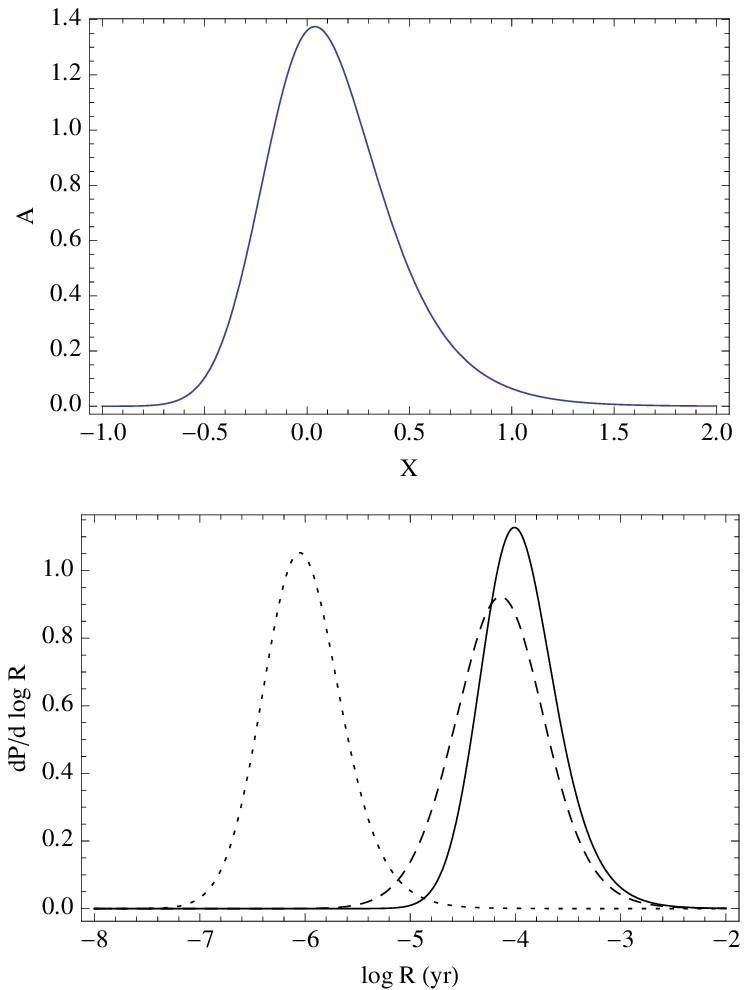}
\caption{\label{fig:GlobalAmbiguity}Top panel: Rate ambiguity function ${\cal A}$ implied by the scaling relations and luminosity
  model uncertainties listed in Eq. (\ref{eq:AmbiguityInputs}), shown versus $X$  (a dimensionless variable corresponding
  to logarithmic rate differences $\log R/R_o$).
Bottom panel: our best estimates for the birthrate of tight PSR-NS (solid), wide PSR-NS (dotted), and tight PSR-WD
binaries (dashed) including marginalization over alternative pulsar luminosity models.  In practice, the curves shown
here correspond to the thick black curves presented previously, after each has been smoothed with the kernel in the top
panel.
These results include the small star formation rate  correction described in \S \ref{sec:sub:lifetime}.
}
\end{figure}

The  above scaling relation and PDF allow us to generalize any result for $\pdf(\log {\cal R}|L_{min},p)$ presented in the main
text, which assumed  a
specific pulsar luminosity model $L_{min,ref}=0.3\unit{mJy}\unit{kpc}^2$ and $p_{ref}=2.0$, to a ``global'' result $p_g(\log {\cal R})$ that fully marginalizes over pulsar model uncertainties:
\begin{eqnarray}
p_g(\log {\cal R})&=& \int \pdf(\log {\cal R}|\log L_{min},p)\pdf(\log L_{min})\pdf(p) d\log L_{min} dp \\
\log Q&\equiv& -\log (L_{min}/L_{min,ref}) + 1.6(p-p_{ref})  \\
\pdf(\log {\cal R}|L_{min},p)&=& \pdf(\log {\cal R} - Q | L_{min,ref}, p_{ref}) \\
\end{eqnarray}
Therefore, changing variables to $z=1.6(p-p_{ref})$, we find the fully-marginalized  PDF $\pdf_g(\log {\cal R})$ follows from the results
presented here via convolution with an ambiguity function ${\cal A}$:
\begin{eqnarray} 
\pdf_g(\log {\cal R})&=& \pdf(\log {\cal R}|L_{min,ref}, p_{ref}) * \pdf(-\log L_{min}) * \pdf(z) \\
{\cal A}&\equiv&  \pdf(-\log L_{min}) * \pdf(z)
\end{eqnarray}
Fig.\ \ref{fig:GlobalAmbiguity} shows our estimate of the global ambiguity function, given
the model of Eq. (\ref{eq:AmbiguityInputs}).   This distribution is roughly as wide as the distributions shown in the text, whose width is limited by Poisson statistics of the number of observed binaries.  Thus, without a better resolved luminosity model, even several new binary pulsars would not reduce the overall uncertainty in the rate estimate. 

\subsection{Lognormal versus Power-law}
\begin{figure}
\includegraphics{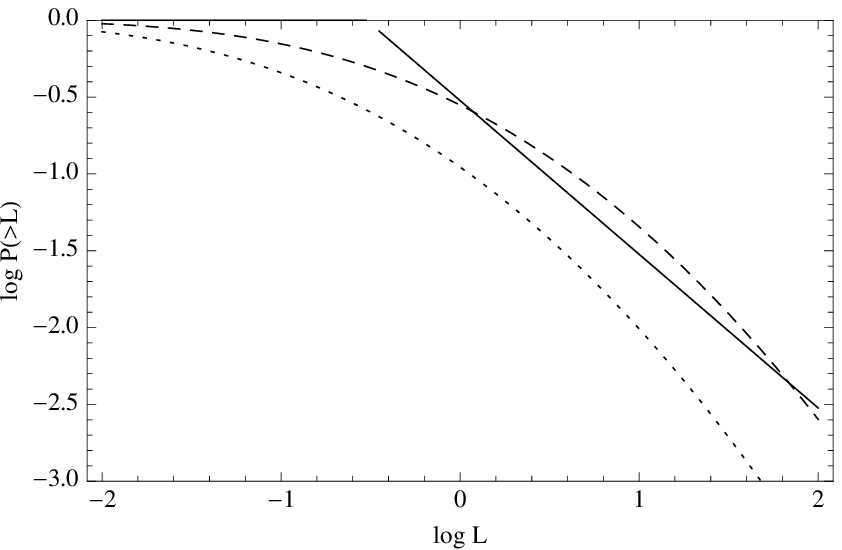}
\caption{\label{fig:Luminosity:Cumulatives}Cumulative luminosity distributions $P(>L)$ for (i) the canonical power-law (solid); (ii) a lognormal distribution with
$\sigma_{\log L}=0.9$ and $\left< \log L \right> = -1.1$ (dotted); and (iii) a lognormal distribution with $\left<\log L \right>  = \log
L_{\rm min}$ where $L_{\rm min}=0.3 \unit{mJy}\unit{kpc}^{-2}$ is our fidicual cutoff for the power law distribution (dashed). }
\end{figure}

Though simple, a global power-law luminosity distribution leads to predictions that depend sensitively on the
low-luminosity cutoff; see Eq. \ref{eq:AmbiguityInputs}.  Recent detailed pulsar population synthesis studies by 
\citet{2006ApJ...643..332F} (henceforth FK) suggest a simpler, lognormal luminosity function fits the whole pulsar population
\begin{eqnarray}
p(\log L)d \log L &=& \frac{1}{\sqrt{2\pi \sigma_{\log L}^2}}e^{-(\log L -<\log L> )^2/2\sigma_{\log L}^2} d\log L
\end{eqnarray}
where  $\sigma_{\log L} =0.9$; see their Fig 15.  
As shown in Figure \ref{fig:Luminosity:Cumulatives}, however, a naive lognormal distribution differs substantially from
our canonical model: even adopting $\left< \log L \right>=\log 0.3\, (\unit{mJy}\unit{kpc}^{-2})$, substantially larger than the best-fit model in
FK with $\left< \log L \right>=-1.1 $, a lognormal distribution predicts more bright and fewer faint pulsars than our fiducial
power-law luminosity model.
Translating to $N_{psr}$ and birthrates, compared to our fiducial power-law distribution  roughly four to six times  more binaries  must be present in
the fiducial lognormal ($\left< \log L \right> = -1.1$) for one to be detected, depending on the pulsar spin and width
being simulated.   For comparison, for each pulsar binary our ``extreme''  lognormal
distribution ($\left< \log L \right>=\log 0.3$) predicts roughly the same numbers $N_{psr}$ of pulsars as the
fiducial power law distribution.\footnote{The lognormal distribution predicts almost all pulsars will be faint and
  therefore visible only from a certain characteristic distance.  The standard power-law distribution, weighted against
  disk area, allows a few bright pulsars to be seen far away (i.e., $P(>L)r^2 \propto 1$).  Thus even though the
  cumulative $P(>L)$ for our ad-hoc ``extreme'' lognormal is always below the fiducial powerlaw, they produce comparable
$N_{psr}$.}
Further, the best fit lognormal
  distribution from FK is really a \emph{spin-dependent} luminosity $L\propto P^{\epsilon_P}\dot{P}^{\epsilon_{\dot{P}}}$, with
  weakly constrained exponents  $\epsilon_{\dot{P}}$ and $\epsilon_{P}$.  That model implicitly introduces 
  time-dependent selection effects, beyond the scope of this paper.   The logormal model described above is presented as a
  convenient summary, not a fundamental distribution; FK thus don't describe how reliable its parameters are.
Lacking control over the model, we defer a detailed discussion of lognormal luminosity models to a future paper.










\optional{
\clearpage
\section{TO DO LIST}
\input{util:todo.tex}
}

\end{document}

%% file: tab-manual-KnownPulsars-PSRNS-deluxetable.tex
\begin{deluxetable}{llllllllllllrrc}[t]
\tablewidth{0pc}
\tabletypesize{\footnotesize}
\tablecaption{Properties of PSR-NS binaries considered in this work. \label{tab:psrns}
{\rm For most pulsars, $f_{b,eff}$ averages over
  the half-opening angle $\rho$ and misalignment angle $\alpha$. For PSR J0737-3039B, we adopt the preferred choice for
  $\alpha \simeq 90^o$ and average only over the stated uncertainties in $\rho$(P$_{\rm s}$). For PSRs B1913+16 and
  B1534+12, where both $\alpha,\rho$ measurements are available, we adopt the values of $f_{b,obs}$ from
  \citet{knst}. The final column is $C= \tau_i/N_{psr}f_{b}$, also see, Eq.\ (\ref{eq:Posterior:Rate:KnownGeometry}) and
  Eq. (\ref{eq:def:taui}).  When numbers are uncertain, this table shows  self-consistent fiducial choices.
  Significant uncertainties are included by explicit convolutions described in the text.   Small uncertainties are
  ignored; for example, our   Monte Carlo estimates for  $N_{psr}$ 
have  poisson sample-size errors of roughly $1/\sqrt{N_{det}}\simeq O(2-5\%)$, where $N_{det}=10^6/N_{psr}$.
}
}
\tablehead{
\colhead{PSR Name} & \colhead{$P_{\rm s}$} & \colhead{$\dot{P_{\rm s}}$} & \colhead{$M_{\rm psr}$} & \colhead{$M_{\rm c}$} & \colhead{$P_{\rm orb}$} & \colhead{e} & \colhead{$f_{\rm {b,obs}}$} & \colhead{$f_{\rm {b,eff}}$} & \colhead{$\tau_{\rm age}^{a}$} & \colhead{$\tau_{\rm mgr}$} & \colhead{$\tau_{\rm d}$} & \colhead{$N_{\rm psr}$} & \colhead{$C$} & \colhead{Ref$^{b}$} \\
\colhead{} & \colhead{(ms)} & \colhead{\tiny{$10^{-18}$} (ss$^{-1}$)} & \colhead{($M_\odot$)} & \colhead{($M_\odot$)} & \colhead{(hr)} &\colhead{}  &\colhead{}   &\colhead{}    & \colhead{(Gyr)} & \colhead{(Gyr)} & \colhead{(Gyr)} & \colhead{}  & \colhead{(kyr)} & \colhead{}\\ 
\cline{1-15}
}

\startdata
tight binaries & & & & & & & & & & & & & & \\

 {B1913+16} & 59. & 8.63 & 1.44 & 1.39    & 7.75 & 0.617 & 5.72 & 2.26 & 0.0653 & 0.301 & 4.31 & 576 & 111 \optional{{\bf 93.6} 95.5} & 1,2 \\
 {B1534+12} & 37.9 & 2.43 & 1.33 & 1.35   & 10.1 & 0.274 & 6.04 & 1.89 & 0.200 & 2.73 & 9.48 & 429 &  1130 \optional{{\bf 1002.3} 1020.} & 3,4\\
 {J0737-3039A}& 22.7 & 1.74 & 1.34 & 1.25 & 2.45 & 0.088 &      & 1.55  & 0.142 & 0.086 & 14.2 & 1403 & 105
 \optional{108 calculated {\bf 90.4} 93.8} & 5\\
 {J0737-3039B}&  2770. & 892. &  &        & 2.45 & 0.088 &      & 14. & 0.0493  & & 0.039 &  &  & 6\\

 {J1756-2251} & 28.5 & 1.02 & 1.4 & 1.18 & 7.67  & 0.181 &  & 1.68 & 0.382 & 1.65 & 16.1 & 664 &  1821 \optional{1828 calculated{\bf 2307.5 } 2440} & 7\\
 {J1906+0746} & 144. & 20300. & 1.25 & 1.37 & 3.98 & 0.085 &  & 3.37 & 0.000112  & 0.308 & 0.082 & 192 & 126 \optional{{\bf 78.2} 77} & 8,9 \\

wide binaries & & & & & & & & & & & & & & \\

 {J1518+4904} & 40.94 & 0.028& 1.56 & 1.05 & 206.4 & 0.249    &     & 1.94  & 29.2 &  $>\tau_{H}$ & 51.0 & 276 & 18,700 \optional{{\bf 17730.5} 18100} & 10, 11\\
 {J1811-1736} & 104.18& 0.901& 1.60 & 1.00 & 451.2 & 0.828    &    & 2.92   & 1.75 &  $>\tau_{H}$ & 7.9 & 584 & 5860 \optional{{\bf 5687.2} 5560} & 12,13\\
 {J1829+2456} & 41.01 & 0.053& 1.14 & 1.36 & 28.3  & 0.139    &     & 1.94  & 12.3 & $>\tau_{H}$ & 43.0 & 271 & 19,000 \optional{{\bf 19349.8} 19000} & 14\\
 {J1753-2240}$^{c}$ & 95.14 & 0.97 & {\it 1.25} & {\it 1.25} & 327.3 & 0.303   & &   2.80      & 1.4  & $>\tau_{H}$   & 8.2 & 270 &  13,900 \optional{{\bf 5974.6 } 5880} & 15\\
\enddata
\tablenotetext{a}{Whenever available, we use the spin-down ages corrected for the Shklovskii effects given in \citet{KT09agecorrection}. As for PSRs B1913+16 and B1534+12, we adapt the results from \citep{spindown}. For PSRs J1906+0746, J1811-1736, J1829+2456, J1753-2240, which are not mentioned in \citet{KT09agecorrection}, we adopt the characteristic age as the current age of the pulsar.}
\tablenotetext{b}{References:
(1) Hulse \& Talor (1975);\nocite{B1913discovery}
(2) Wex, Kalogera, \& Kramer (2001);\nocite{B1913kick}
(3) Wolszczan (1991);\nocite{B1534discovery}
(4) Stairs et al.\ (2002);\nocite{B1534timing}
(5) Burgay et al.\ (2003);\nocite{J0737Adiscovery}
(6) Lyne et al.\ (2004);\nocite{J0737Bdiscovery}
(7) Faulkner et al.\ (2004);\nocite{J1756discovery}
(8) Lorimer et al.\ (2006);\nocite{J1906discovery}
(9) Kasian, and PALFA consortium (2008);\nocite{J1906update}
(10) Nice, Sayer, \& Taylor (1996);\nocite{J1518discovery}
(11) Janssen et al.\ (2008);\nocite{J1518update}
(12) Lyne et al. (2000);\nocite{J1811discovery}
(13) Kramer et al.\ (2003);\nocite{J1811update}
(14) Champion et al. (2004);\nocite{J1824discovery}
(15) Keith et al.\ (2008)\nocite{J1753discovery}
}
\tablenotetext{c}{The nature of the companion of PSR J1753-2240 is not yet clear, and it can be either a WD or NS \citep{J1753discovery}. In this work, we assume PSR J1753-2240 is another wide NS-NS binary. Given that its small contribution to the total rate estimates, we note that the nature of the companion would not change the main results shown in this work. The masses shown for PSR J1753-2240 are half the total binary mass.  All plausible mass pair choices lead to a merger time $>10\unit{Gyr}$; the masses otherwise do not influence our results.} 
\end{deluxetable}

%% file: tab-manual-KnownPulsars-PSRWD-deluxetable.tex
\begin{deluxetable}{lllllllllllrrc}[t]
\tablewidth{0pc}
\tabletypesize{\footnotesize}
\tablecaption{ Properties of tight PSR-WD binaries considered in this work.  \label{tab:psrwd}
{\rm For all binaries here, $f_{\rm {b,eff}}$ averages over the half-opening angle $\rho$ and misalignment angle
  $\alpha$.Monte Carlo sampling uncertainty in $N_{psr}$ is roughly $1/\sqrt{N_{det}}\simeq O(3-5\%)$
}
}
\tablehead{
\colhead{PSR Name} & \colhead{$P_{\rm s}$} & \colhead{$\dot{P_{\rm s}}$} & \colhead{$M_{\rm psr}$} & \colhead{$M_{\rm c}$} & \colhead{$P_{\rm orb}$} & \colhead{e} & \colhead{$f_{\rm {b,eff}}$} & \colhead{$\tau_{\rm age}^{a}$} & \colhead{$\tau_{\rm mgr}$} & \colhead{$\tau_{\rm d}$} & \colhead{$N_{\rm psr}$} & \colhead{$C$} & \colhead{Ref$^{b}$} \\
\colhead{} & \colhead{(ms)} & \colhead{\tiny{$10^{-18}$} (ss$^{-1}$)} & \colhead{($M_\odot$)} & \colhead{($M_\odot$)} & \colhead{(hr)} &\colhead{}   &\colhead{}    & \colhead{(Gyr)} & \colhead{(Gyr)} & \colhead{(Gyr)} & \colhead{}  & \colhead{(kyr)} & \colhead{}\\
\cline{1-14}
}
\startdata
 {J0751+1807} & 3.48 & 0.00779 & 1.26 & 0.12 & 6.32  & $<10^{-7}$      & 2.62  & 6.66  & 9.48 \optional{1.4}  & $>\tau_{H}$  & 2404 & 1588 \optional{{\bf 409.9} 508} & 1,2\\
 {J1757-5322} & 8.87 & 0.0278 & 1.35 & 0.67 & 10.9   & $<10^{-6}$      & 1.26  & 7.16  & 8.0  & 145    & 1082 & 7335 \optional{{\bf 9412.1} 6050} & 3\\
 {J1141-6545} & 393.9 & 4295. & 1.3 & 0.986 & 4.74   & 0.172           & 5.46  & 0.00145 & 0.60 & 0.10 & 346 &  53 \optional{
   {\bf 53.83} 55}   & 4,5\\
 {J1738+0333} & 5.85 & 0.0241 & 1.7 & 0.2 & 8.5      & $4\times10^{-6}$& 1.69  & 3.71  & 10.8 & $>\tau_{H}$  & 609 & 9716 \optional{{\bf 6223.5} 6140} & 6 \\
%
\enddata
\tablenotetext{a}{For PSR J0751+1807, we use the spin-down ages corrected for the Shklovskii effects \citep{KT09agecorrection}. For other pulsars, we use the characteristic age.} 
\tablenotetext{b}{Reference:
(1) Lundgren, Zepka, \& Cordes (1995);\nocite{J0751discovery} (2) Nice, Stairs, \& Kasian (2008) ;\nocite{J0751update}
(3) Edwards \& Bailes (2001) ;\nocite{J1757discovery}
(4) Kaspi et al.\ (2000) ;\nocite{J1141discovery}
(5) Bailes et al.\ (2003) ;\nocite{J1141update}
(6) Jacoby (2005)\nocite{J1738discovery}
}

\end{deluxetable}

%% file: tab-mma-KnownPulsars-MisalignmentOptions-CKrevised.tex
\begin{deluxetable}{llllll}[t]
\tablewidth{40pc}
\tabletypesize{\small}
\tablecaption{Comparison \label{tab:Conclusions} of effective beaming correction factors $\fbeff$ predicted by different pulsar geometry models (columns 3--6): flat ($p(\alpha)=2/\pi$ ; adopted in the text, see Eq.\ \ref{eq:alphaflat}); $p(\alpha)=\cos\alpha$; $p(\alpha)$ as proposed by Eq.\ (\ref{eq:pdf:Observed}) of \citet{2003PASJ...55..461Z}  (ZJM03); and $p(\alpha)\propto 1/\alpha$ between $3^\circ$ and $90^\circ$.   Reading from left to right, misalignment models correspond to an increasing fraction of fairly aligned pulsars; for example, the last distribution has half of all binaries with $\alpha<16^\circ$, which for the pulsars of interest usually corresponds to  $f_{\rm b}>10$.  Nonetheless, the relevant effective beaming correction factor does not increase as rapidly as the largest (or even median) $f_{\rm b}$.  Generally, when $p(\alpha)$ has a subpopulation of  tightly-aligned pulsars, $\fbeff$ is inversely proportional to the fraction of binaries with \emph{large} misalignments, as well as being  directly proportional to the median $f_{\rm b}$ for the large-misalignment subpopulation.  Therefore, large $\fbeff$ values are only possible if a significant fraction of pulsars are assumed to be tightly aligned; such a  hypothesis is difficult to test, as these tightly aligned pulsars
  will be extremely difficult to see. 
}
\tablehead{
\colhead{Name} & \colhead{$f_{b,obs}$} & \colhead{flat} & \colhead{ZJM03}  & \colhead{$\cos \alpha $}  & \colhead{$1/\alpha$} 
}
\startdata
  tight PSR-NS &  &  &  &  &\\ 
 {B1913+16} & 5.72 & 2.26 & 2.63 & 2.62 & 3.37 \\
 {B1534+12} & 6.04 & 1.89 & 2.14 & 2.13 & 2.69 \\
 {J0737-3039A} &  & 1.55 & 1.70 & 1.69 & 2.08 \\
 {J1756-2251} &  & 1.68 & 1.88 & 1.87 & 2.33 \\
 {J1906+0746} &  & 3.37 & 4.05 & 3.98 & 5.25 \\
\hline
 wide PSR-NS &  &  &  &  &\\ 
 {J1518+4904} &  & 1.94 & 2.21 & 2.21 & 2.81 \\
 {J1811-1736} &  & 2.92 & 3.46 & 3.42 & 4.48 \\
 {J1829+2456} &  & 1.94 & 2.22 & 2.21 & 2.79 \\
 {J1753-2240} &  & 2.8 & 3.3 & 3.27 & 4.28 \\
\hline
 tight PSR-WD &  &  &  &  &\\ 
 {J0751+1807} &  & 2.62 & 3.08 & 3.06 & 4.00 \\
 {J1757-5322} &  & 1.26 & 1.35 & 1.33 & 1.57 \\
 {J1141-6545} &  & 5.46 & 6.66 & 6.46 & 8.65 \\
 {J1738+0333} &  & 1.69 & 1.90 & 1.89 & 2.37 
\end{deluxetable}